\documentclass[%
 aip,
 amsmath,amssymb,
 reprint,%
]{revtex4-1}

\usepackage{graphicx}
\usepackage{dcolumn}
\usepackage{bm}

\usepackage[utf8]{inputenc}
\usepackage[T1]{fontenc}
\usepackage{mathptmx}
\usepackage{etoolbox}
\usepackage[bottom]{footmisc}

\usepackage{url}

\makeatletter
\def\@email#1#2{%
 \endgroup
 \patchcmd{\titleblock@produce}
  {\frontmatter@RRAPformat}
  {\frontmatter@RRAPformat{\produce@RRAP{*#1\href{mailto:#2}{#2}}}\frontmatter@RRAPformat}
  {}{}
}%
\makeatother
\begin{document}

\preprint{AIP/123-QED}
\title{Reinforcement Learning for Photonic Component Design}
\author{Donald Witt*}
 \affiliation{Department of Electrical and Computer Engineering, The University of British Columbia, V6T-1Z4, Vancouver, British Columbia, Canada\\}
  \affiliation{Stewart Blusson Quantum Matter Institute, V6T 1Z4, Vancouver, British Columbia, Canada\\}
  \email[Corresponding Author: ]{donald.witt@alumni.ubc.ca} 

\author{Jeff Young}%
 \affiliation{Stewart Blusson Quantum Matter Institute, V6T 1Z4, Vancouver, British Columbia, Canada\\}
 \affiliation{ 
Department of Physics and Astronomy, The University of British Columbia, V6T 1Z1, Vancouver, British Columbia, Canada\\
}%

\author{Lukas Chrostowski}
 \affiliation{Department of Electrical and Computer Engineering, The University of British Columbia, V6T-1Z4, Vancouver, British Columbia, Canada\\}
  \affiliation{Stewart Blusson Quantum Matter Institute, V6T 1Z4, Vancouver, British Columbia, Canada\\}
 
\date{28 May 2023}



\begin{abstract}
We present a new fab-in-the-loop reinforcement learning algorithm for the design of nano-photonic components that accounts for the imperfections present in nanofabrication processes. As a demonstration of the potential of this technique, we apply it to the design of photonic crystal grating couplers fabricated on an air clad 220 nm silicon on insulator single etch platform. This fab-in-the-loop algorithm improves the insertion loss from 8.8 to 3.24 dB. The widest bandwidth designs produced using our fab-in-the-loop algorithm can cover a 150 nm bandwidth with less than 10.2 dB of loss at their lowest point. 
\end{abstract}

\maketitle

\section{Introduction}
The design of photonic components by traditional methods is a complex, time and labor intensive process. This process starts with a theoretical model. Simulations are then run to verify this model and obtain more accurate performance estimates. This process is computationally intensive, and thus, it takes a long time to obtain accurate results. Once simulations are complete and a final design is chosen, the components must then be manually optimized for the fabrication process. The challenge in doing so is that despite efforts to model the effects of lithography,\cite{Litho1,Litho2,NvidiaLitho} the models only consider process bias and smoothing, thus resulting in a mismatch between the simulated geometry and the fabricated one. 
The complexity of this design process has limited the wide scale adaptation of integrated photonics as it requires domain specific knowledge, limiting the pool of potential component designers. 

Machine learning has shown that computers are able to analyze data and come up with valuable insights in many fields. Deep learning is a subset of machine learning where an artificial neural network is used to learn from a dataset and provide these insights. Deep learning has been applied to many fields with great success, such as image classification,\cite{originalimage,Batchnormalization,Resnet} anomaly detection,\cite{deepanomaly} medical imaging,\cite{cardiacml,medicalcomparison} and photonics.\cite{deeplearningphotonics} Reinforcement learning (RL) is a subset of deep learning that uses feedback in the form of a reward or score from an environment to generate new, improved actions. RL has been shown to come up with creative solutions to problems in games,\cite{openaihideandseek,aieconomist} applications in robotic control,\cite{robotcube,robotwalking} materials,\cite{RLmaterials} and component design.\cite{RLphotonics,RLphotonics2,Nvidia} 

Prior machine learning approaches to component design rely on simulations to obtain their datasets, resulting in the same two disadvantages found in traditional design methods. First, the simulation of photonic components takes an extremely long time to obtain accurate results. Second, simulations, even corrected with lithography models, do not account for all imperfections present in every nanofabrication process.

In this paper, we show that we can successfully overcome the challenge of achieving optimal device performance by inventing a new approach: fab-in-the-loop reinforcement learning. This approach incorporates feedback from measurements of prior fabricated designs to produce new, improved designs using reinforcement learning. We demonstrate that by applying fab-in-the-loop reinforcement learning to photonic component design, we are able to produce components with better performance than we are able to produce using traditional design methods. These best performing designs are largely unintuitive and would be unlikely to be suggested by an expert designer. 
Without detailed knowledge of the fabrication deviations, it would be fortuitous if human-generated design would outperform our best performing designs.

\begin{figure*}
     \includegraphics[width=1.0\textwidth]{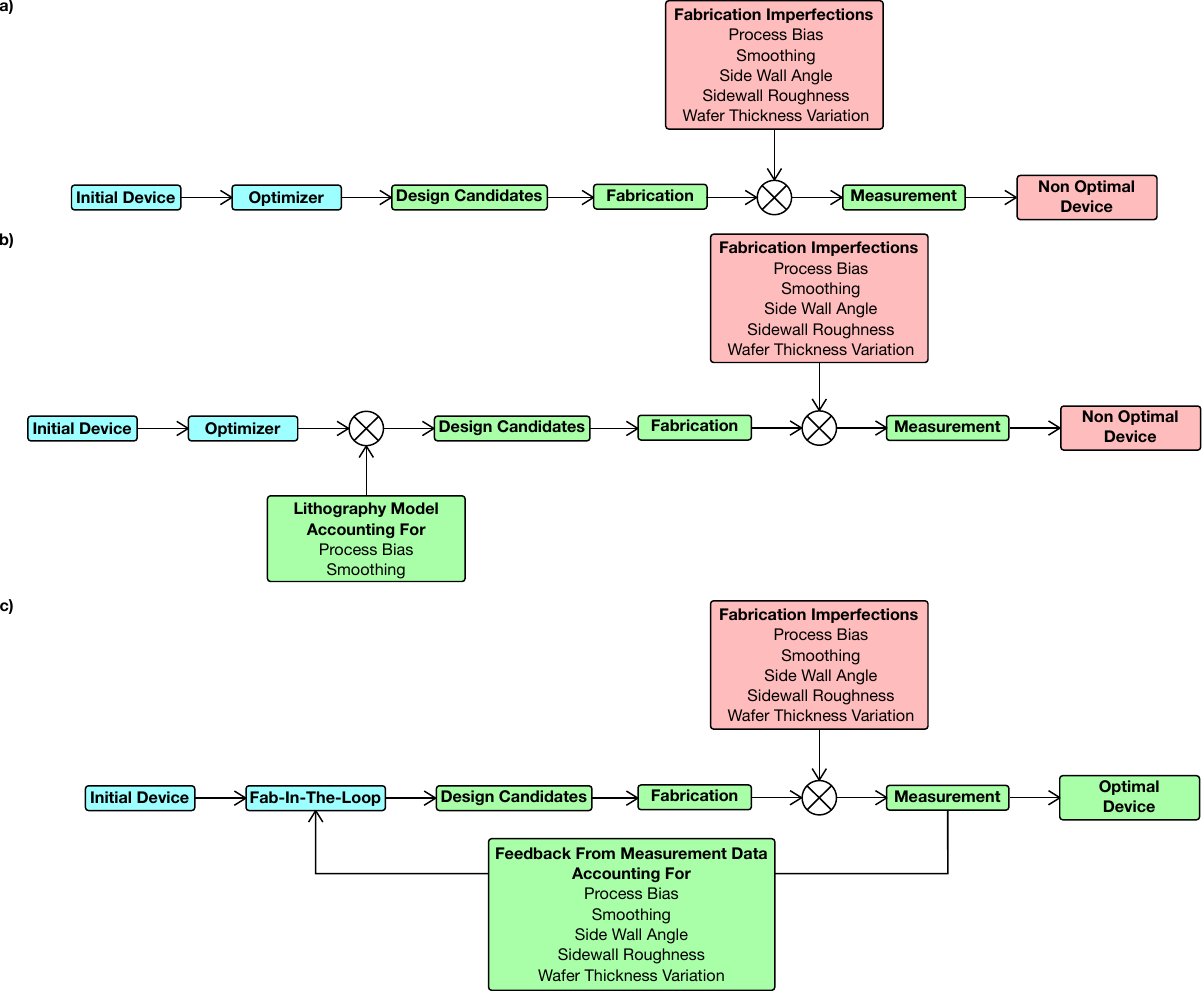}
      \caption{A comparison between traditional device optimization techniques vs the fab-in-the-loop approach. a) In the traditional approach, the optimizer will produce a design based on simulation results. The user will then fabricate this design, measure it, and find the performance drastically different from that predicted by the simulation due to various fabrication effects. b) The user introduces a lithography model to correct for the process bias and smoothing, but this model will not account for other fabrication effects. c) In the fab-in-the-loop approach, the algorithm will automatically optimize the device to the fabrication process without additional user input based solely on the measured results.}
       \label{traditional_vs_fab_in_loop}
\end{figure*}

Our approach is also efficient. To obtain a new dataset using fab-in-the-loop takes approximately one week with our in-house fabrication and measurement capabilities: Training and running the fab-in-the loop RL algorithm takes a day on an older Mac laptop with an 8-core Intel i9 processor. Fabrication of the new chip takes 1.5 days. Automated measurement takes 3 days.  If the data contained on one chip were to be obtained using traditional simulation approaches and available simulation hardware, it would take approximately 2 years using 3D Finite-Difference Time-Domain (FDTD) simulations on a server equipped with dual Intel Xeon Gold 5118 processors  and 96 GB of RAM without accounting for fabrication effects. Using RL to generate improved devices is, therefore, only possible with our fab-in-the-loop technique Fig. \ref{traditional_vs_fab_in_loop}.

To demonstrate the power of our technique, we apply fab-in-the-loop RL to grating coupler design. Optical coupling to and from an integrated photonic circuit with low loss and high bandwidth is critical for many applications. Grating couplers are a convenient method to achieve this as they allow for the rapid testing of photonic devices. They do so by surface coupling light into waveguides, which allows for them to be placed anywhere within a design, in contrast to edge coupling. Used in conjunction with automated measurement setups, they allow for the rapid characterization of thousands of individual photonic circuits across a chip. Unfortunately, grating couplers are typically plagued by high insertion loss and/or narrow operating bandwidths. They are also sensitive to fabrication process variations. These issues have traditionally required knowledgeable individuals to design custom grating couplers to support different fabrication processes and operating wavelengths as required for classical and quantum applications.\cite{quantumReview,Tcenter,Wcenter}

Utilizing fab-in-the-loop RL, we obtained a grating coupler design with an insertion loss of 3.24 dB as compared with an insertion loss of 8.8 dB for a design that utilizes the traditional design methodology on our single etch air clad silicon on insulator (SOI) platform. The widest bandwidth designs produced with our technique cover the 150 nm bandwidth of our laser with less than 10.2 dB of loss at their lowest point. With these results, we can choose an optimal grating coupler design for each application.
 
In section 2, we present the parameterized grating coupler design used as a demonstration of our fab-in-the-loop RL. In section 3, we describe the fab-in-the-loop RL algorithm. In section 4, we discuss the device optimization process. Finally, in section 5, we discuss the results.

\section{The Parameterized Grating Coupler}
\begin{figure}[!h]
     \includegraphics[width=0.9\columnwidth]{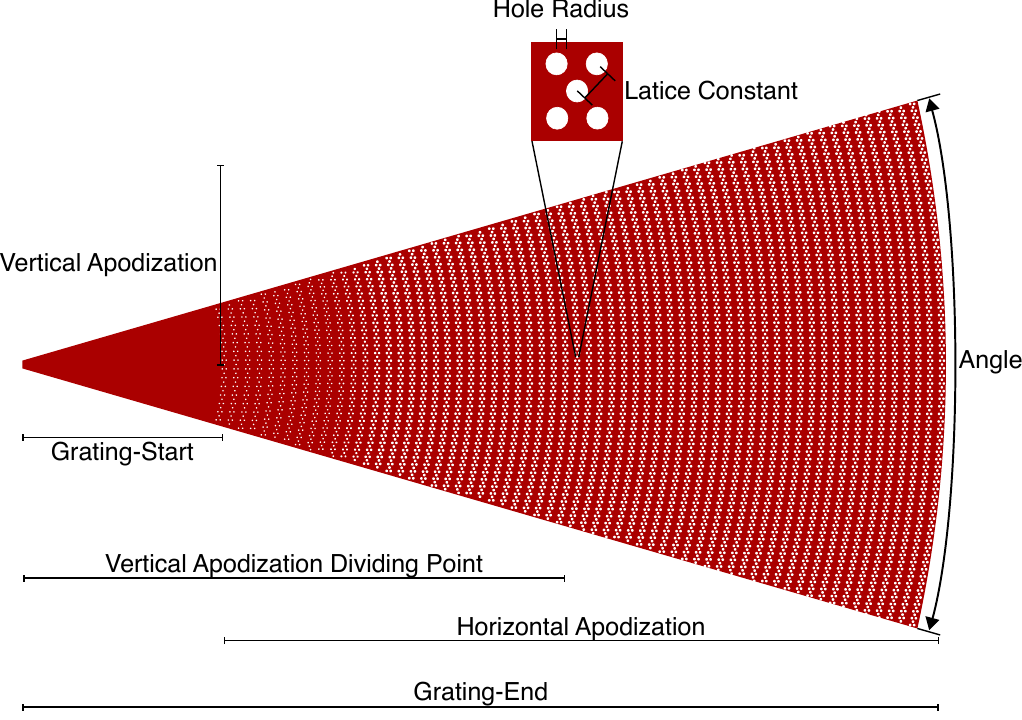}
      \caption{A schematic of the parameterized grating coupler. The start of the grating, end of the grating, angle, hole radius and lattice constant are all adjustable. The horizontal apodization start and end are adjustable. The vertical apodization adjusts the hole radius, as you move out from the center line. The vertical apodization dividing point allows for two different values of this parameter. Not shown here is the hole diameter at which the lattice constant is adjusted instead of the hole radius. In total, there are 12 adjustable parameters.}
       \label{gc_gds}
\end{figure}

A parameterized grating coupler design was created as an initial starting point for the fab-in-the-loop RL. This is key to the process as a parameterized design reduces the search space required, resulting in faster convergence of the neural network results and, thus, requiring fewer training rounds. This design is illustrated in Fig. \ref{gc_gds}. It is parameterized by 12 geometric quantities with ranges given in Table \ref{table:parametertable}. This initial design was developed to conform to the limits imposed by our fabrication process as described below. The ranges of the 12 parameters given in Table \ref{table:parametertable} have been chosen to have a wide search space while still maintaining a compact device footprint. 

Our fabrication process is based on a 525 nm thick positive electron beam resist Zep520A.\cite{SiEPICFab} This places some constraints on the design. The maximum aspect ratio of Zep520A is 5:1; this means that traditional sub-wavelength designs\cite{NRC_gc,subwave} are not possible with our process due to the resist collapsing when fabricating the sub-100 nm isolated features required for such designs. To overcome this challenge, a design based on sub-wavelength holes instead of sub-wavelength lines is used. The lattice design is based on the design from L. Liu et al.\cite{normphcgc} To reduce the device area, focusing has been added to this traditional photonic crystal grating coupler (PhCGC) design. We also include focus-angle variation, and both horizontal and lattice apodization. The purpose of the horizontal apodization is to improve the insertion loss and bandwidth as has been shown by Y. Ding et al.\cite{phcgcap} This apodization is defined from its starting point with the hole diameter decreasing until it reaches the minimum hole diameter, at which point it will remain fixed until it reaches its endpoint. The vertical apodization is to improve the focusing of the light into the waveguide. This parameter allows for the hole diameter to be increased or decreased moving out from the center. Two different values of this apodization can be used as specified by its dividing point. This is useful for fine tuning the focus of the fiber spot into the waveguide. Both these types of apodizations are affected by another parameter, the hole diameter at which to switch to lattice apodization. This means that once the hole diameter is at or below this level, the lattice will be expanded instead of the hole diameter being reduced further. This is done as both the spacing of the holes and the hole diameter affect the effective index. However, the minimum hole diameter is affected by the fabrication process. By increasing the lattice constant, one can obtain a similar effect to a smaller hole size without the fabrication limit. The angle is the focus angle of the grating coupler cone. The grating-start defines where, on the grating coupler, the holes start to be drawn and the grating-end defines where the grating coupler ends. Table \ref{table:parametertable} enumerates all 12 parameters and their ranges. The range of these parameters has been chosen to have a wide search space while still maintaining a compact device footprint. This parameterized design is then optimized by the fab-in-the-loop RL algorithm.

\begin{table}
\caption{PhCGC Parameters and Ranges}
\centering
\begin{tabular}{ | m{5cm} | m{1.5cm}| m{1.5cm} | } 
\hline
 \textbf{Parameter} &  \textbf{Minimum Value}&  \textbf{Maximum Value}\\ 
 \hline
 Initial Hole Radius & 35 nm & 200 nm \\  
 \hline
 Lattice Constant  & 35 nm & 500 nm \\    
 \hline
 Grating-End  & 80 holes & 250 holes \\   
 \hline
 Grating-Start  & 0 holes & 70 holes \\   
 \hline
 Horizontal Apodization Start  & 0  holes & 70 holes \\   
 \hline
 Horizontal Apodization End  & 20 holes & 250 holes \\   
 \hline
 Minimum Hole Diameter  & 10 nm & 100 nm \\   
 \hline
 Hole Diameter for Lattice  Apodization & 50 nm & 150 nm \\   
 \hline
 Vertical Apodization Dividing Point & 10 holes & 150 holes \\   
 \hline
 Vertical Apodization Before Dividing Point & -0.1 & 0.1 \\   
 \hline
 Vertical Apodization After Dividing Point & -0.1 & 0.1 \\   
 \hline
 Angle  & 10° & 45° \\   
 \hline
\end{tabular}
 \label{table:parametertable}
\end{table}

\section{The Reinforcement Learning Algorithm}
Fabricated devices are needed to provide data to our fab-in-the-loop RL algorithm. An initial design is input as the starting point to our fab-in-the-loop algorithm, which then generates a set of 1250 photonic components, which have a range of design parameters. In our case, our initial input is the best traditionally optimized design for each wavelength bin in wavelength range of 1490-1640 nm based on the available laser. In addition, four different control grating couplers with hole radius of 80, 100, 120, and 140 nm respectively, are added to the chip for determination of the current process bias.
We fabricate a chip with these components, measure their spectra, and generate a dataset consisting of the spectra and design parameters for each one.

 This dataset is then input back into our fab-in-the-loop RL algorithm, which then generates new, better performing designs. Now the initial input is the best measured design from the previous round for each wavelength bin in wavelength range of 1490-1640 nm. A new chip with the new designs is made, and the process is repeated until an optimal design as needed for the particular application is achieved.

An issue with our approach is that the measured dataset will still be smaller than in traditional applications of RL. To overcome this defect, our fab-in-the-loop RL algorithm consists of both a spectral predictor and a traditional deep deterministic policy gradient algorithm (DDPG).\cite{ddpgpaper,ddpgcode} The spectral predictor generates estimated power spectra from photonic device design parameters after training on previously measured devices. 
The DDPG algorithm proposes new designs based on a score determined from the parameters and the spectra provided by the spectral predictor. 

\begin{figure}[!h]
     \includegraphics[width=0.9\columnwidth]{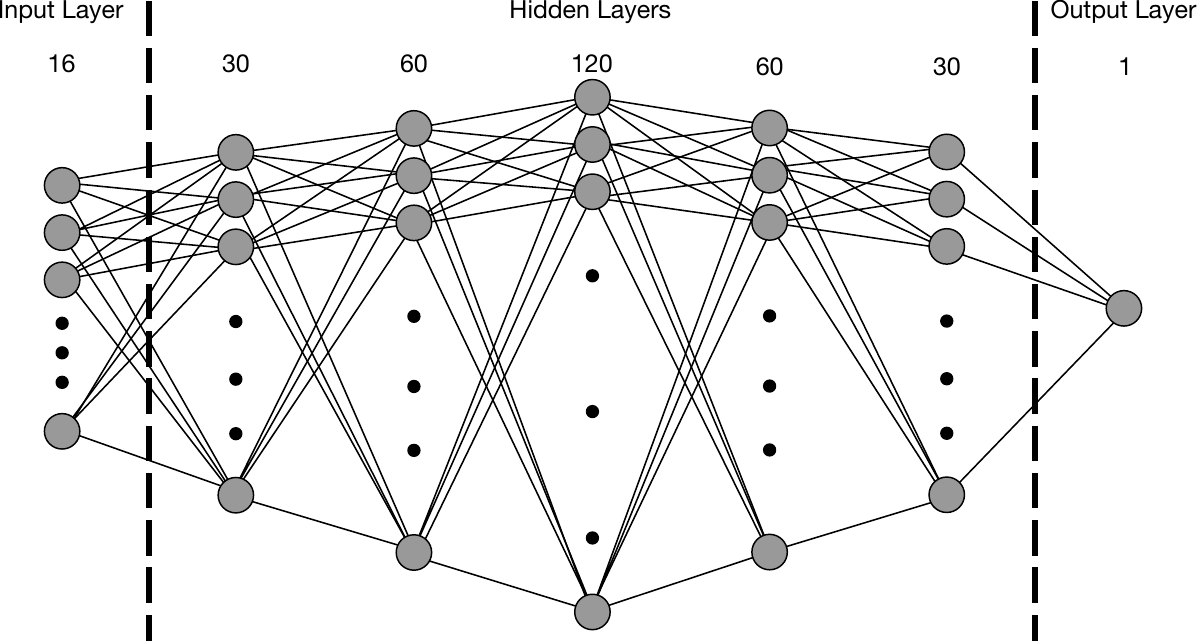}
      \caption{The spectral predictor. This network takes the 12 parameters of the grating coupler design and the current process bias for 80, 100, 120, and 140 nm holes and produces a power estimate for one wavelength. One hundred and fifty copies of this network are used to produce a power spectrum consisting of 150 wavelength values between 1490 and 1640 nm.}
       \label{spectral_device_figure}
\end{figure}

The spectral predictor consists of a neural network as described in Fig. \ref{spectral_device_figure} that estimates the device spectrum based on the input parameters. 

The spectral predictor is trained from measured data. First, a chip is measured after a fabrication round. The current process bias is determined from the optical spectrum of the set of four different control grating couplers with hole radius of 80 nm, 100 nm, 120 nm and 140 nm respectively. This is done by calculating the correlation function of the optical spectrum of the design with the same hole radius with varying bias steps and the spectrum for the same design with no bias from the initial dataset. The bias which results in the maximum correlation with the original unbiased result is then determined to be the current process bias for that hole radius. This allows for any changes in the fabrication process to be accounted for.

The spectral data from the chip is then processed to remove any invalid measurements due to detector saturation, incorrect number of data points or other setup related errors during the measurement process. Then the data for any devices with a minimum insertion loss of greater than 40 dB is dropped. The reason for this is twofold: It is possible that the device was damaged during the fabrication or measurement process and could still be potentially a good design. It would negatively effect the algorithm if these designs were given falsely low scores due to a flawed measurement. The second reason is that the initial rounds are likely to fabricate large numbers of poorly performing designs. Inclusion of a large number of poorly performing designs can cause the spectral predictor to produce poor quality predictions for all designs. 

A mean squared error loss function is utilized to train the spectral predictor with a learning rate of 0.0001 on 10 000 examples drawn at random from the above dataset.\cite{backprop} 

Now that the spectral predictor is trained, the DDPG algorithm generates a new set of designs. The DDPG portion of the algorithm consists of identically dimensioned actor-critic networks with two layers of size $600$ and $400$, respectively. The input is 14 parameters, consisting of the 12 parameters of the grating coupler design and two parameters specifying the desired operating wavelength bin. The output is 12 improved grating coupler parameters. The DDPG proposes a new set of parameters. The spectral predictor generates a spectrum.

A score is then computed, which is provided to the DDPG. The score is determined, using the principle of reward shaping.

If all 12 device parameters are in range, then the score is calculated using the estimated power spectrum found by the spectral predictor. 
The average power is calculated over the target wavelength range and then this power is normalized.
The score $S$ is 
\begin{equation} 
\label{inrange_score}
S=\begin{cases}
10^{(1+np)}& r<a\\
10^{(1+np-\frac ra)}& r>a
\end{cases}
\end{equation}
where $np$ is the normalized power, $r$ is the hole radius, and $a$ is the lattice constant. 
The score penalizes designs in which the hole radius $r$ is greater than the lattice constant $a$ as they generally perform poorly. 

If one or more of the device parameters $p_i$ is out of range, first, a parameter score is calculated for each out of range parameter $p_i$:
\begin{equation}
\label{outofrange_score}
 ps_i=\begin{cases}
-1-2\big|\frac{p_i}{\max(p_i)}\big|& p_i>\max(p_i)\\
-1-2\big|\frac{\min(p_i)}{p_i}\big|& p_i<\min(p_i)
\end{cases}
\end{equation}
where $\max(p_i)$ and $\min(p_i)$ are the maximum and minimum allowed values for that parameter. The score is then given by
\begin{equation} 
\label{outofrange_score_final}
S=10^{\frac{ \sum_i ps_i}{22}}
\end{equation}
where the sum is over the set of out of range parameters. Again, this score penalizes out of parameter range designs.

\begin{figure}[!h]
     \includegraphics[width=0.9\columnwidth]{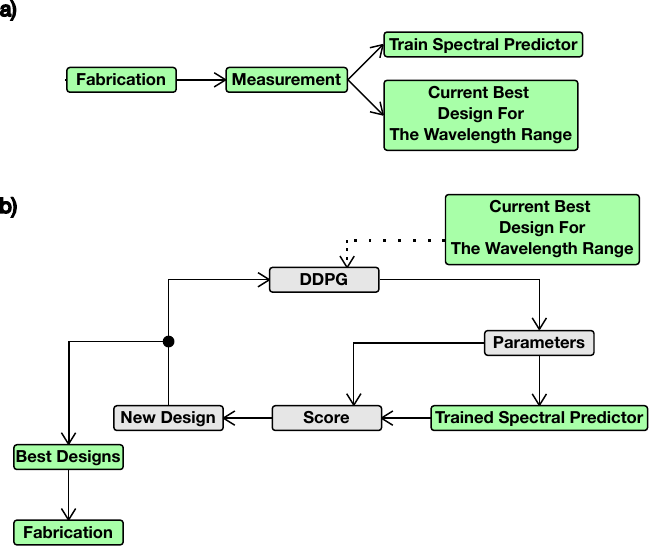}
      \caption{A schematic of our fab-in-the-loop RL algorithm.  a) The feedback from measurement data. The measurement data are used to both train the spectral predictor and update the current best design. b) The DDPG portion of the algorithm that produces new designs. A DDPG episode is started with the current best design. Then, the DDPG algorithm produces a new set of design parameters. These parameters are then scored using the scoring algorithm described in Eqs. \ref{inrange_score}-\ref{outofrange_score_final}. This score is combined with the parameters and fed back into the DDPG algorithm until the end of the training episode. This is repeated 10 000 times for each of the required wavelength ranges.}
       \label{algorithm}
\end{figure}

The score calculated in Eqs. (\ref{inrange_score} - \ref{outofrange_score_final}) is used by DDPG\cite{ddpgpaper,ddpgcode} to generate new designs as seen in Fig. \ref{algorithm}b.

  For training, the variables describing the training process are $\alpha= 0.000 005$, $\beta= 0.0005$ and $\tau=0.0001$, and a batch size of $32$ was used.\cite{ddpgpaper} $10 000$ training episodes were used for each fabrication run. A training episode is one round of the DDPG algorithm for each of the requested wavelength ranges. The length of the training episodes, that is, the number of attempts at improving the design, is determined in the following way: If all parameters of the device are in range, a length of 10 is used. Otherwise, after 3 out of range actions, the episode is ended early. New designs with all parameters in range are saved.  The top 1250 designs are fabricated in the next cycle.

After fabrication, the new devices are measured, and a new dataset is generated. The new data are then used to train the spectral predictor and update the current best devices used to generate improved designs. 

\begin{figure}[!h]
     \includegraphics[width=0.9\columnwidth]{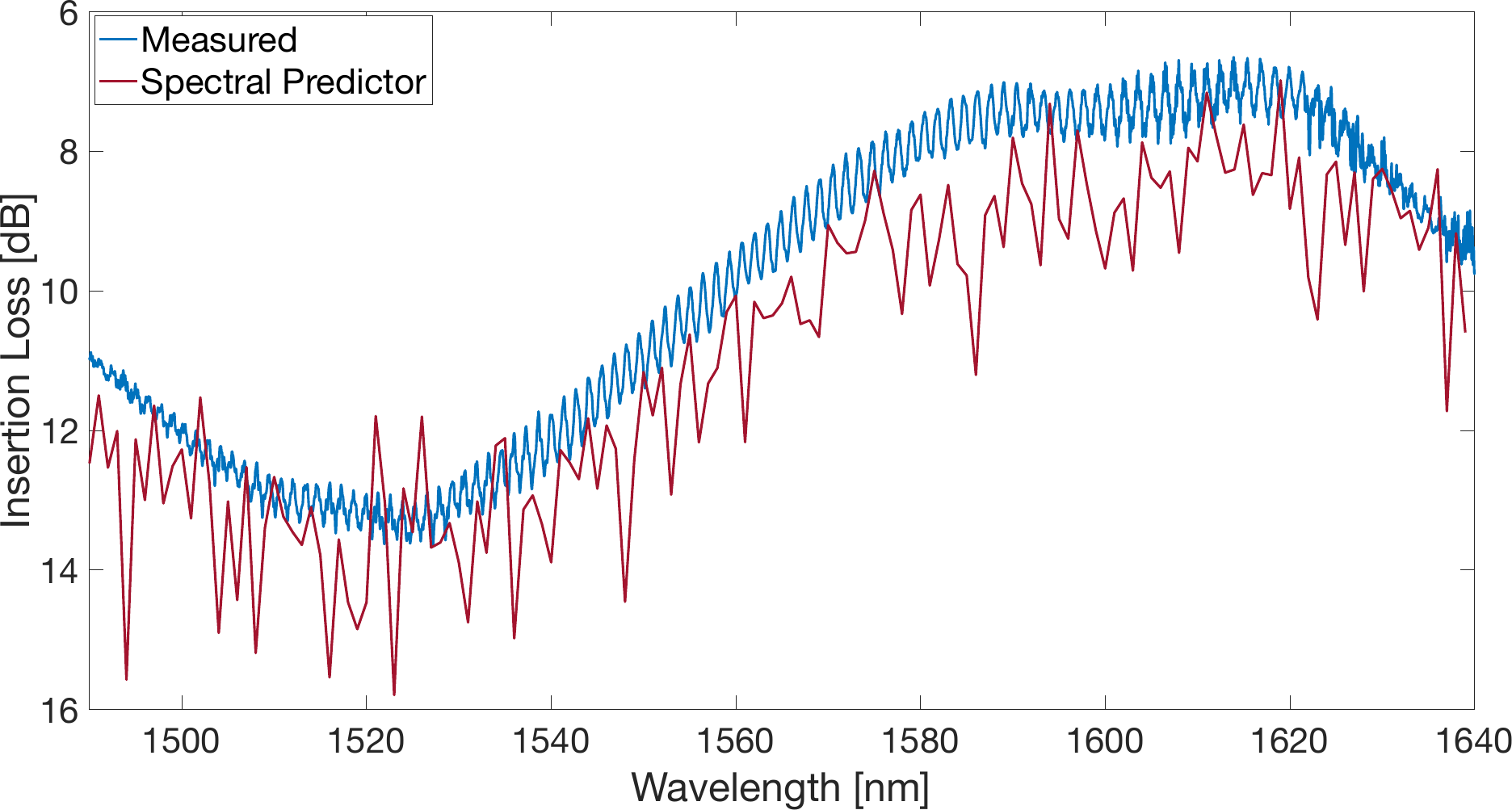}
      \caption{An example of the output from the spectral predictor in red compared with a measured spectrum in blue. It can be seen that they match well in this case.}
       \label{predict_plot}
\end{figure}

\section{Optimization}

The fab-in-the-loop cycle is started. 10 000 episodes of the DDPG algorithm are run for 10 requested wavelengths bins between 1490 and 1640 nm. These wavelength bins are kept constant after being selected so the fab-in-the-loop RL algorithm returns improved designs for each wavelength. A new chip is then fabricated. This chip contains 1250 examples selected from the best scoring designs from the algorithm training run. For calibration and training purposes, the current top five fabricated designs in each wavelength bin, along with four different random biases between -20 and +20 nm of each of these designs, are added to the chip. This is done to provide the spectral predictor with data on how process bias affects high scoring designs. In addition, the four different control grating couplers with hole radius of 80, 100, 120, and 140 nm respectively are again added to the chip for determination of the current process bias.
The fabricated devices are measured and the dataset is updated. This cycle was repeated six times. 

\begin{figure}[!h]
     \includegraphics[width=0.9\columnwidth]{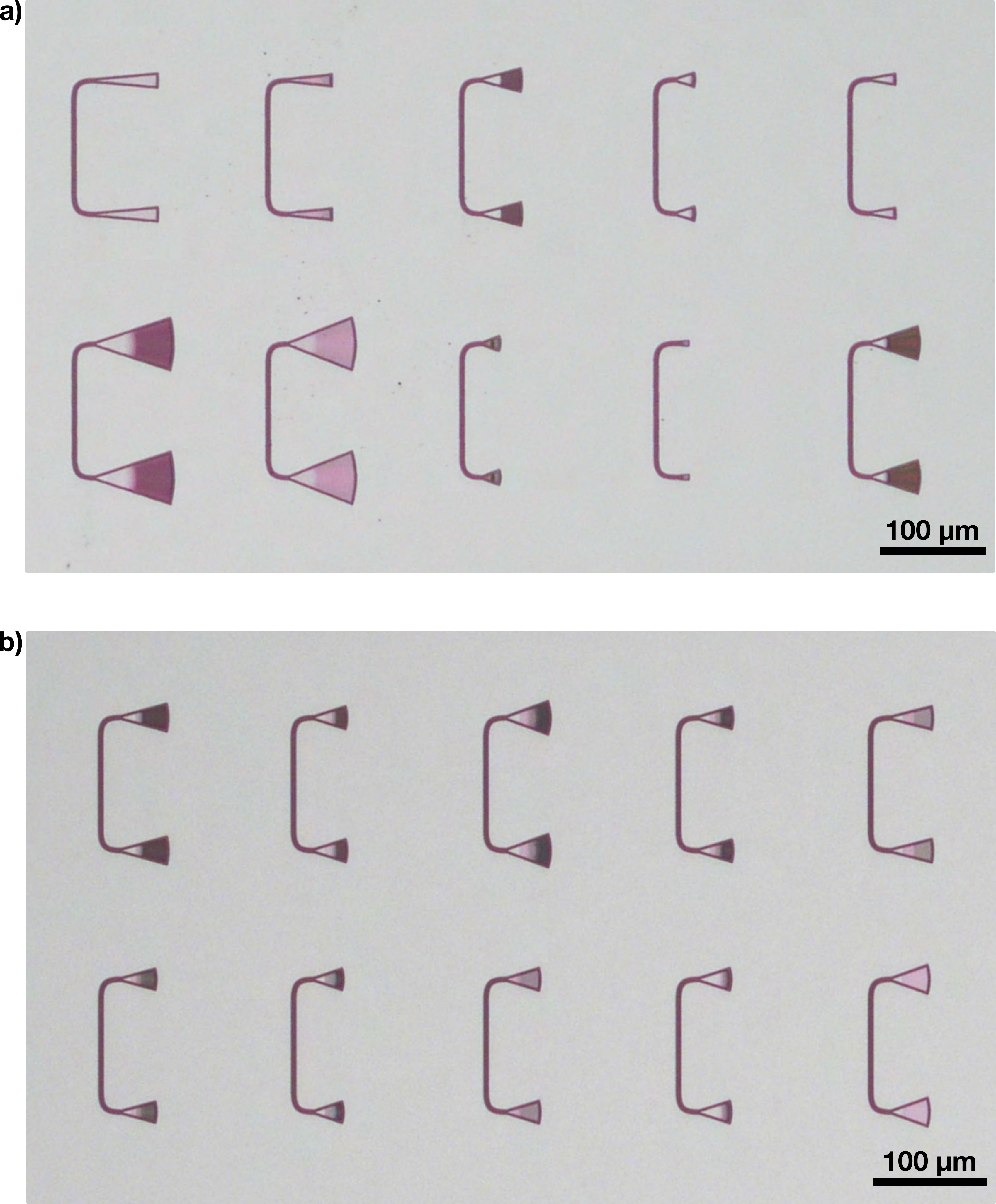}
      \caption{a) An optical microscope image of an area of the chip from the first fabrication run. One can see there are a wide variety of designs with varying parameters. b) An optical microscope image of an area of the chip from the last fabrication run. One can see that the designs have converged around the optimal ones leading to a more constrained parameter variation. }
       \label{example_chip}
\end{figure}

An image of a small area of two chips for the first and last round is displayed in Fig. \ref{example_chip}, which visually exhibits the convergence of the parameters. 

\section{Results}
Six rounds of the fab-in-the-loop RL algorithm were carried out and produced designs that are significantly better than those from traditional design methodology.  The spectral predictor is key to producing these designs.

\begin{figure}[!h]
     \includegraphics[width=0.9\columnwidth]{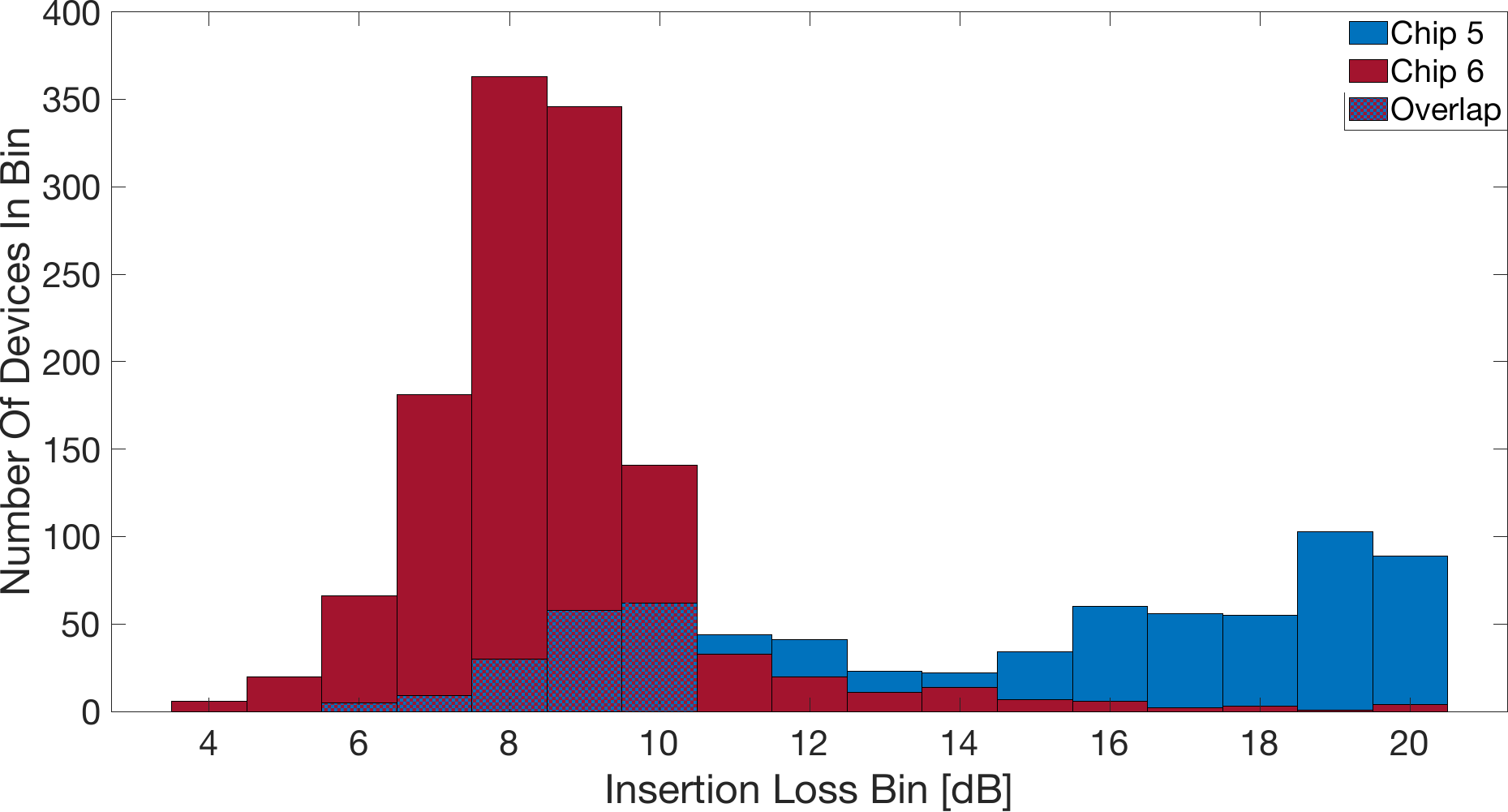}
      \caption{A histogram highlighting the importance of the spectral predictor. Chip 5 is in blue, chip 6 is in red and the overlap between the two is checkered. Round 6 fully utilizes the spectral predictor. With the spectral predictor, the mean has significantly improved from a mean insertion loss of 14.8 dB on chip 5 to a mean insertion loss of 8.6 dB on chip 6. Chip 6 has 766 devices or 61\% of total number of devices with an insertion loss less than 8.8 dB. Chip 5 has only 58 such devices or 4.6\% of the total.}
       \label{chip_to_chip_convergence}
\end{figure}

The importance of the spectral predictor can clearly be seen in Fig. \ref{chip_to_chip_convergence}. A programming error meant that the spectral predictor was not fully utilized to select new designs. This error was corrected for the sixth fabrication run. The result of this can be seen in Fig. \ref{chip_to_chip_convergence}. The spectral predictor successfully eliminates poor designs before fabrication, resulting in a decrease in the mean insertion loss from 14.8 dB on chip 5 - 8.6 dB on chip 6. The mean insertion loss on chip 6 is an improvement over the insertion loss of our initial, traditional design with 8.8 dB of loss. Chip 6 has 1224 out of 1250 designs, which is 98\% of designs, with less than 20 dB of insertion loss. Chip 5 only has 691 out of 1250, which is 55\% of designs, in the same category. Chip 6 has 766 devices with less than 8.8 dB of insertion loss; chip 5 has only 58. The spectral predictor significantly refined the search space around designs that are likely to perform well as highlighted by these results. This significantly improves the chances of finding low loss devices performing well above the mean. 

\begin{figure}[!h]
     \includegraphics[width=0.9\columnwidth]{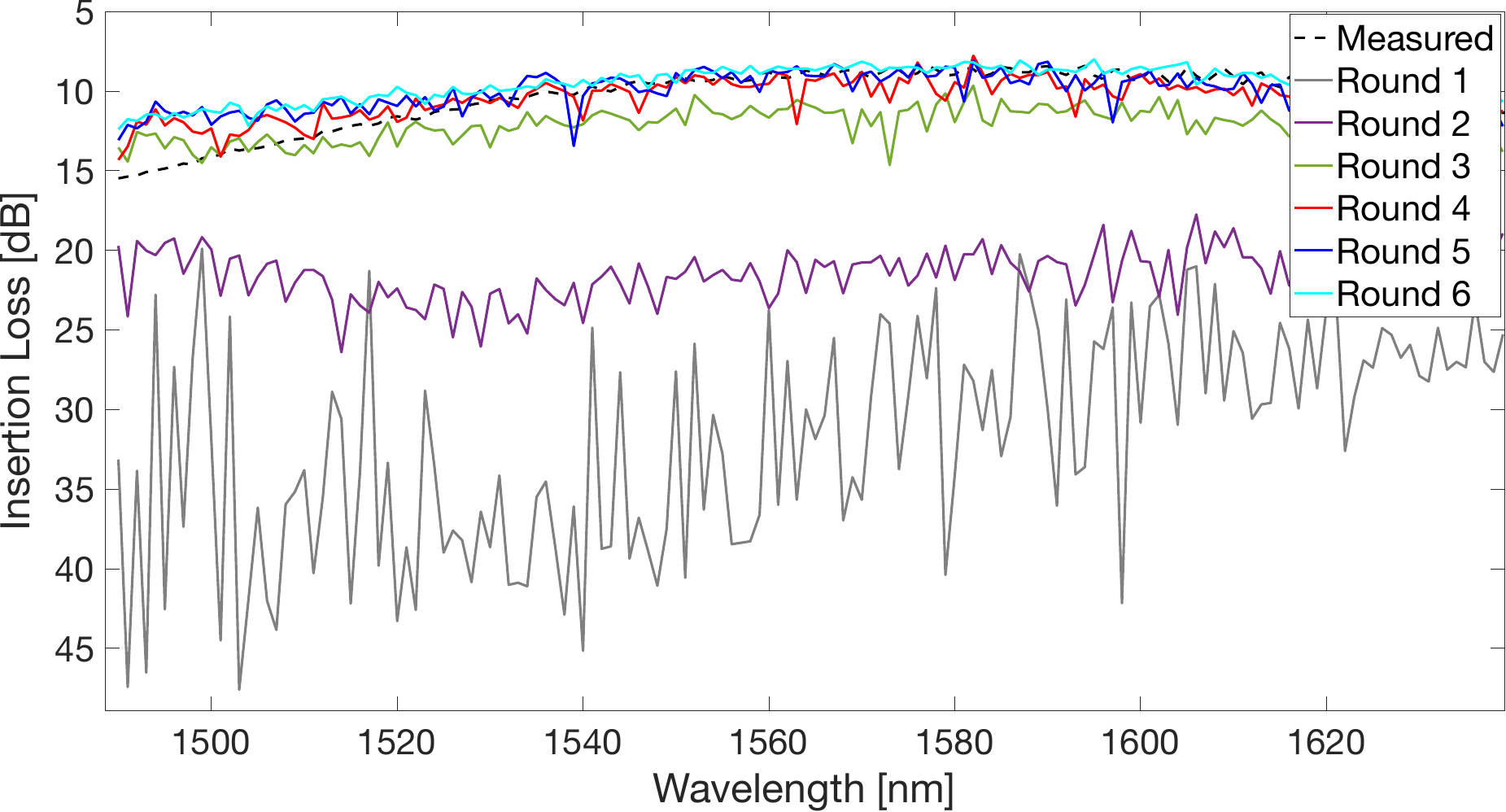}
      \caption{The convergence of the spectral predictor over multiple training rounds for one device. It can be seen that the results are significantly improved after only three training rounds.}
       \label{device_convergence}
\end{figure}

\begin{figure}[!h]
     \includegraphics[width=0.9\columnwidth]{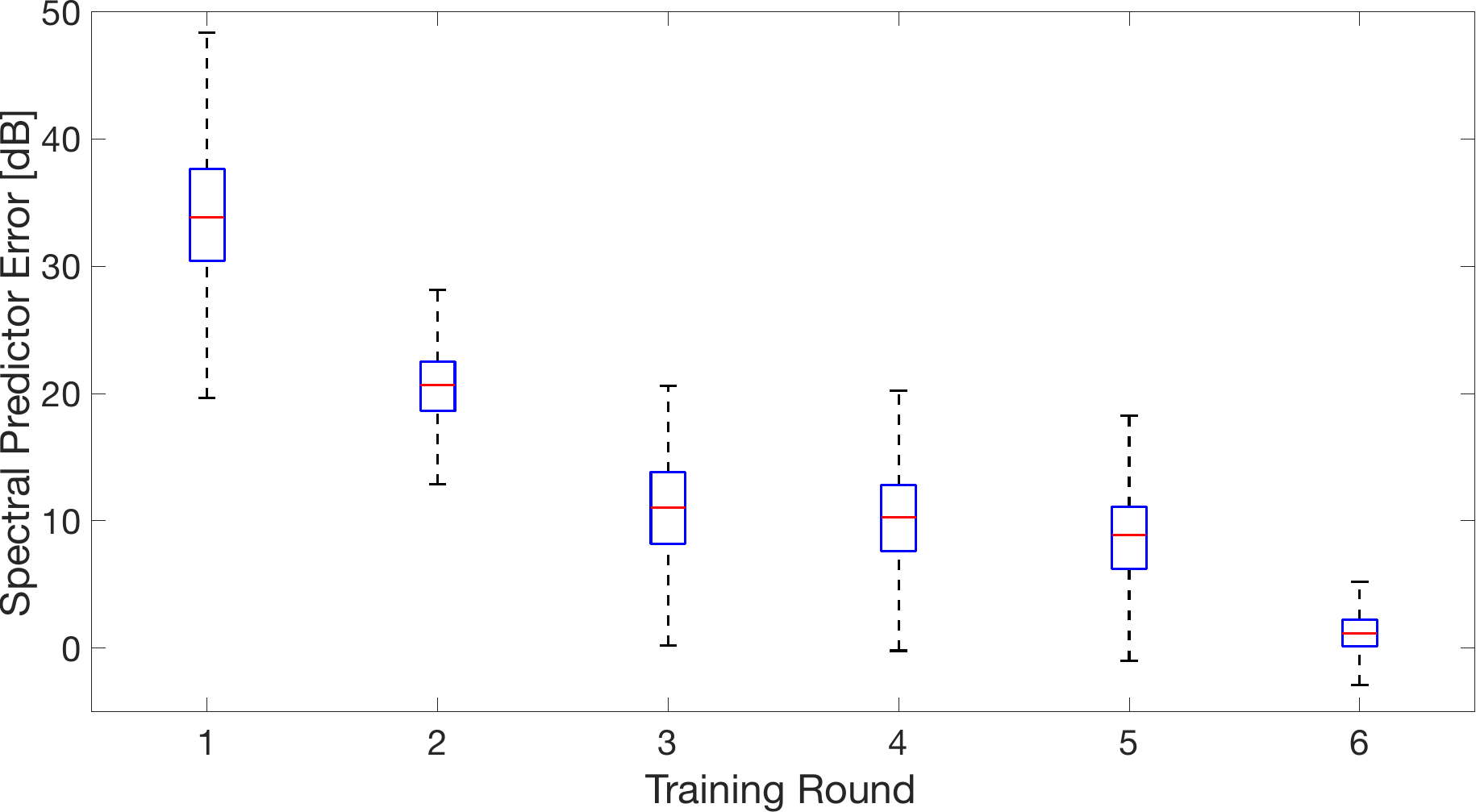}
      \caption{The convergence of the spectral predictor over multiple training rounds. This convergence is calculated using the measurement results from the last fabrication round. Then, the parameters of these devices are fed into the spectral predictor that was saved at various stages of training. The resulting predicted spectrum and measured spectrum are then differenced, and the average error is calculated. One can see that after three training rounds, the error is significantly reduced.}
       \label{overall_convergence}
\end{figure}

The predicted spectrum from the spectral predictor also improved significantly with each fabrication run. Fig. \ref{device_convergence} shows the spectra from one given device as generated by the spectral predictor as trained in rounds 1-6. One sees that the spectrum converges to the measured one. In Fig. \ref{overall_convergence}, we give a box plot of the predicted spectrum from each training round to the measured spectrum from round 6 for the entire set of round 6 devices. Again it can be seen that the error significantly improves after only 3 fabrication runs. 

\begin{figure}[!h]
     \includegraphics[width=0.9\columnwidth]{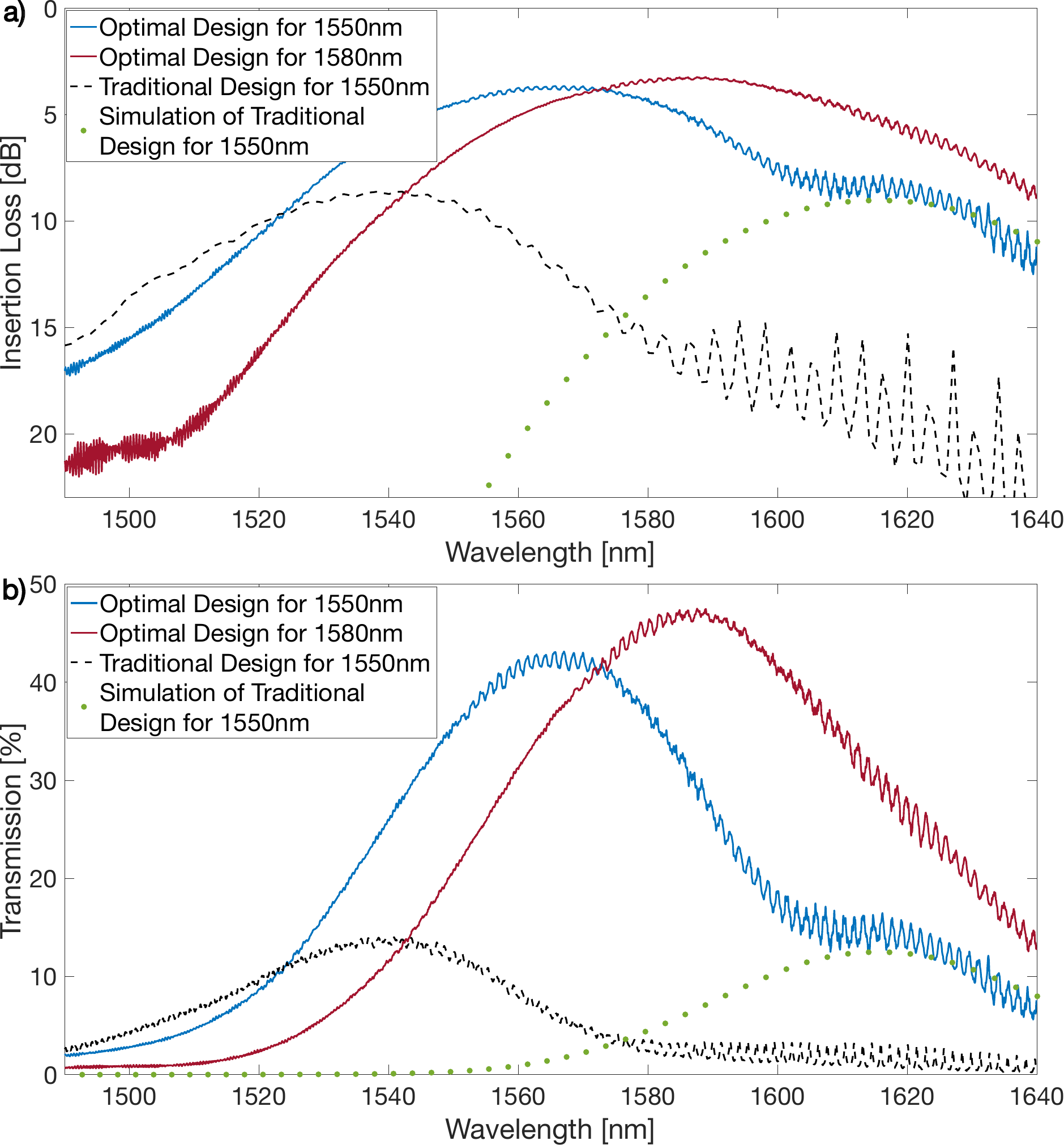}
      \caption{a) A set of PhCGC spectra designed using our algorithm. The spectrum of a traditionally optimized design used as our starting point for the fab-in-the-loop RL algorithm is given by the dashed line. The result of an FDTD simulation of same traditional design is given by the green dotted line. Note that the operating wavelength of the fabricated device is significantly shifted due to fabrication effects. The fab-in-the-loop RL algorithm is able to produce a design at the target wavelength of 1580 nm with an insertion loss of 3.24 dB per coupler. b) The same devices plotted in linear scale. This further highlights the importance of the improvement. With our traditional design, only 14\% of photons make it into or out of the chip. With the optimized design, over 45\% do so. This improvement is critical for quantum applications low photon loss.}
       \label{gc_responses}
\end{figure}

It produced devices optimized for operation at different wavelengths as can be seen in Fig. \ref{gc_responses}. 

\begin{figure}[!h]
     \includegraphics[width=0.9\columnwidth]{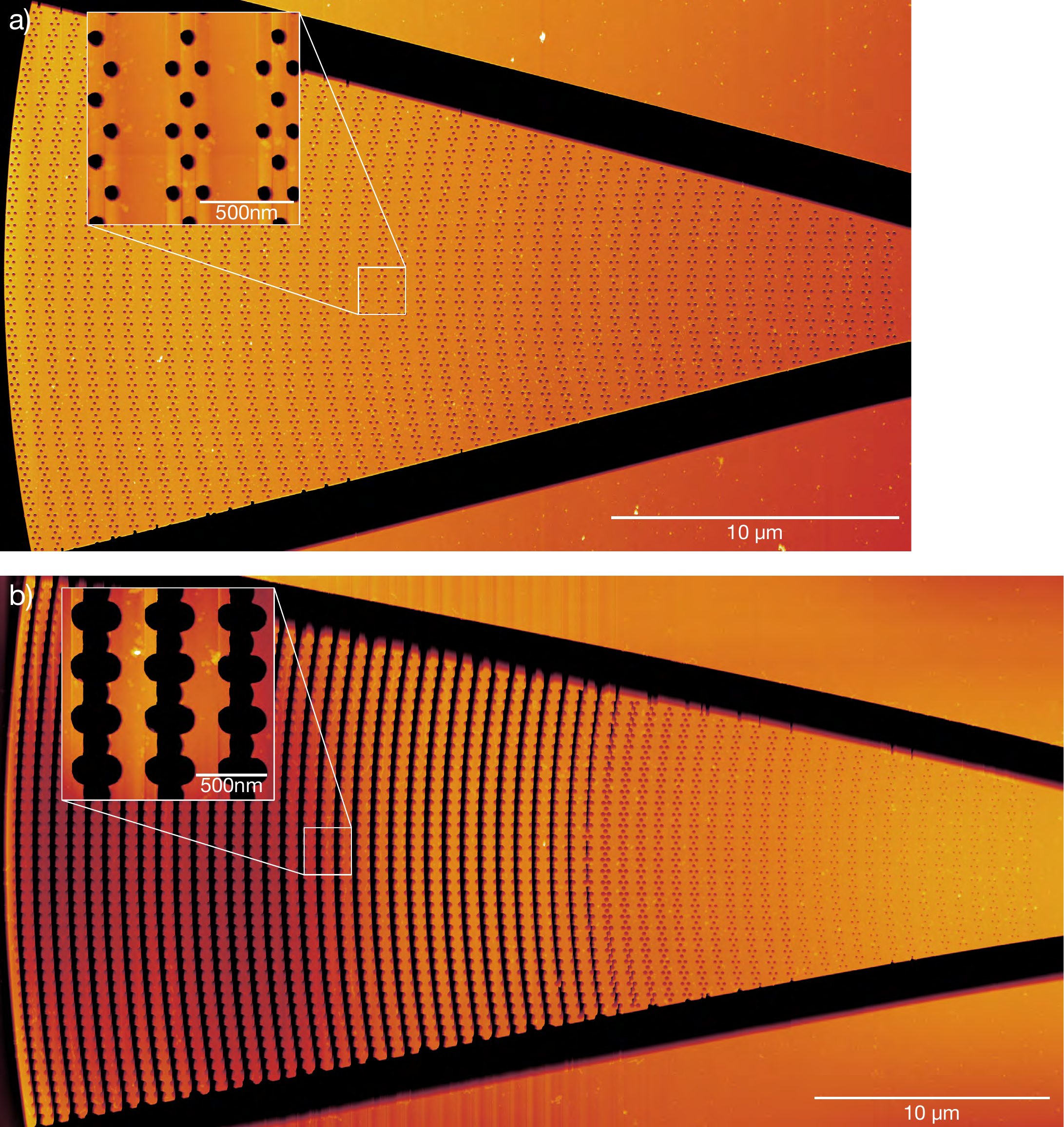}
      \caption{a) An AFM image of a traditional PhCGC. b) A top performing optimized design. This design has an insertion loss of 3.82 dB at 1625 nm. One can see that the design includes corrugations formed by the merging of the holes. The corrugations disappear as the hole size is reduced by the applied apodization. }
       \label{gc_afm}
\end{figure}

The fab-in-the-loop RL algorithm also produced designs with an improved insertion loss of 3.24 dB per coupler as compared to an insertion loss of 8.8 dB for our traditionally optimized design.\cite{SiEPICFab}
An example of such a design is given in Fig. \ref{gc_afm}. This design involves the creation of a combination of lines formed from merged holes and of separated holes. This is interesting as it overcomes the previously mentioned limitation of the original sub-wavelength design from our fabrication process. This design would be unlikely for a human designer to attempt due to the assumption that the merged holes would not work. In addition, it would be difficult to draw in a fashion that could be successfully fabricated. In contrast, the fab-in-the-loop RL algorithm successfully produces several such designs.

\begin{figure}[!h]
     \includegraphics[width=0.9\columnwidth]{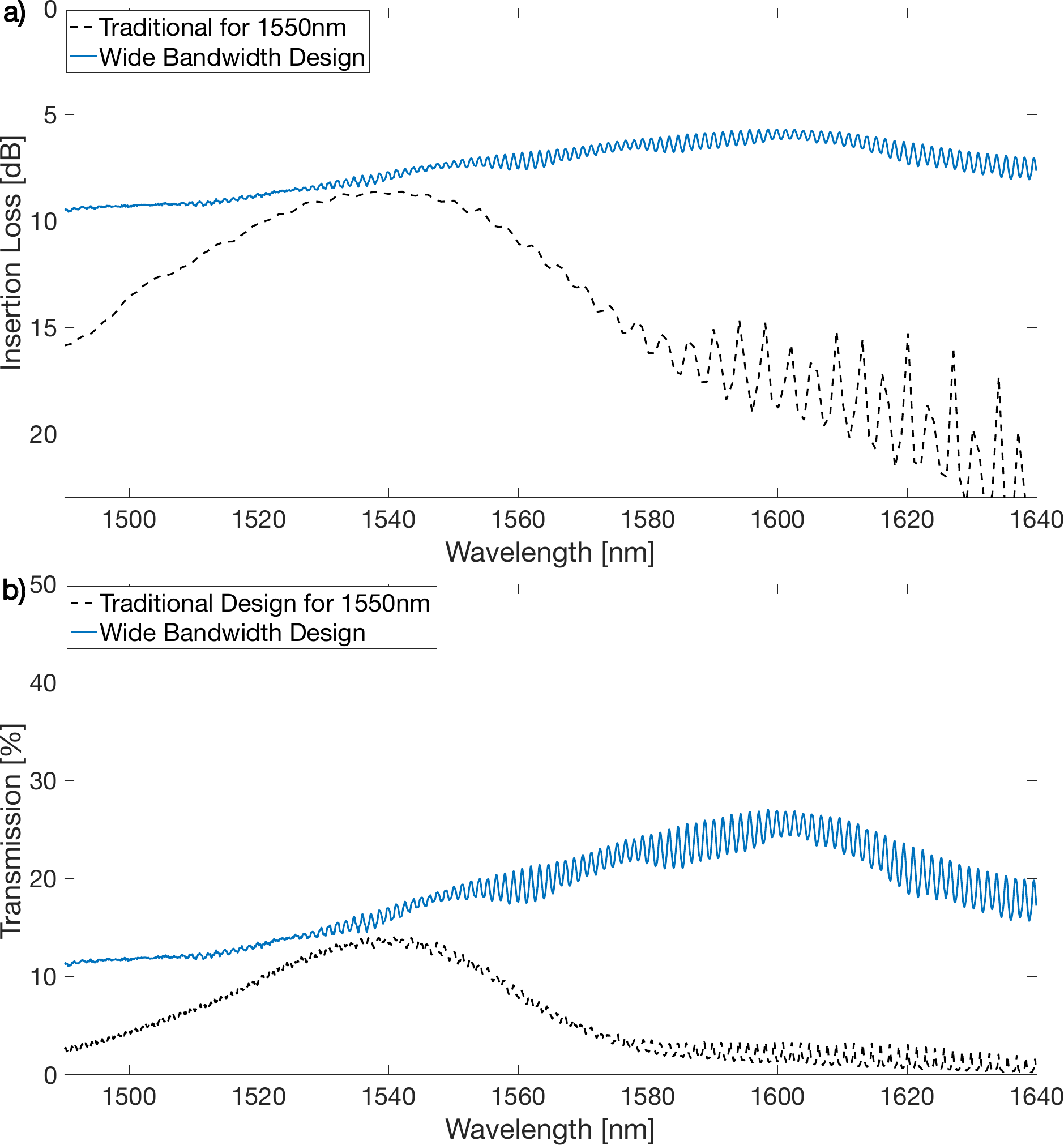}
      \caption{a) A wide bandwidth PhCGC spectrum designed using our algorithm. The spectrum of a traditionally optimized design used as our starting point for the fab-in-the-loop RL algorithm is given by the dashed line. b) The same devices plotted in linear scale. }
       \label{gc_responses_widebandwidth}
\end{figure}

It also produced designs with significantly larger bandwidth than our traditional design. The power spectrum of one such design is given in Fig. \ref{gc_responses_widebandwidth}. This design covers the 150 nm bandwidth of our laser with less than 10.2 dB of loss at its lowest point. Such a design is valuable in the study of other broadband components such as filters.

\section{General Applicability Considerations}
Fab-in-the-loop RL can be applied to other components, such as contra-directional couplers,\cite{contra-directional} splitters,\cite{ybranch} and cavities.\cite{mircavity,nanobeam} Fab-in-the-loop RL could also be applied to the design of other optical devices, such as lasers,\cite{newlaser} and quantum dots\cite{quantumdot} in III-V substrates along with other similar platforms. To do this, one needs a parameterized design and to modify the spectral predictor to provide a prediction of the desired property for the relevant device such as the quality factor, splitting ratio, or filter bandwidth.

For example, fab-in-the-loop RL could be used to optimize a H0 photonic crystal cavity \cite{H0cavity}. 

The first step would be to create a parameterized cell for the cavity. For an H0 cavity, among these would be the radius, lattice constant, the waveguide coupling distance,  shifts in the x direction, and shifts in the y direction necessary to form the cavity. 
The next step would be to determine the desired parameters to optimize, including additional ones important in a fabrication process. 
For example, round holes end up slightly oblong after fabrication. Introducing additional parameters to describe the distortion of the holes would allow this to be taken into account. 
The range of these parameters would be set with consideration to ensure that the device performs as an H0 cavity.

Next, the predictor neural network needs to be implemented. For a cavity, it would be more efficient to predict the central wavelength and quality factor  instead of predicting the entire spectrum. This could be done by training the neural network on these metrics. In the case of a photonic crystal, it is typical to experience 10 nm of variation in the central wavelength for the same design on the same chip due to fabrication variations. On the surface, this would be a challenge for fab-in-the-loop RL. To account for this variation, the central wavelength predictor neural network could be trained using  10 nm bin for the central wavelength.  A set of current process biases would be also input into the spectral predictor to allow it to account for this information.
The fab-in-the-loop RL process could then be initialized with an H0 photonic crystal design and several rounds run to produce optimized designs.

 \section{Conclusion}
The fab-in-the-loop RL process allows for the optimization of nanophotonic components accounting the imperfections present in nano-fabrication process and is able to produce devices with better performance for a given fabrication process. In particular, when applied to grating coupler design, fab-in-the loop produced grating couplers with a measured insertion loss of 3.24 dB per coupler as compared with a measured insertion loss of 8.8 dB for our traditionally optimized design.  It also produces designs optimized to different wavelengths and bandwidths. The widest bandwidth designs produced using our fab-in-the-loop algorithm can cover a 150 nm bandwidth with less than 10.2 dB of loss at their lowest point. As our process utilizes data obtained from devices fabricated with our nano-fabrication process, they are  virtually guaranteed to out-perform conventional brute force, or even fab-uninformed design methods, including other machine learning based techniques. By its nature,  fab-in-the-loop RL process learns from information specific to the fabrication process that fundamentally cannot be fully characterized by other means. We believe that it can be applied to other photonic devices to the same effect.

\section{Acknowledgments}
This work was supported by the Natural Sciences and Engineering Research Council of Canada (NSERC), the B.C. Knowledge Development Fund (BCKDF), the Canada Foundation for Innovation (CFI) and the SiEPICfab consortium. We would also like to thank Dr. Kashif Masud Awan for his initial work on the fabrication process development, Professor Edmond Cretu for the AFM access, and the staff of the Stewart Blusson Quantum Matter Institute’s Advanced Nanofabrication Facility for their assistance. Finally, we would like to thank Professor Kristin Schleich for her encouragement to pursue the project and her helpful advice. 

\section{Author Contributions}
Donald Witt: Conceptualization; Methodology; Investigation; Software; Validation; Data Curation; Writing - Original Draft; Writing - Review \& Editing (Lead); Jeff Young: Writing - Review \& Editing (Support); Resources (Equal); Supervision (Equal); Funding acquisition (Support); Lukas Chrostowski: Writing - Review \& Editing (Support); Supervision (Equal); Resources (Equal); Funding acquisition (Lead).

\section{Data Availability}
Source code is available on GitHub at https://github.com/Donald-Witt/Fab-in-the-loop. Other data supporting the findings is available from the corresponding author upon reasonable request.

\section{References}
\bibliography{reinforcement-learning-photonics}

\providecommand{\noopsort}[1]{}\providecommand{\singleletter}[1]{#1}%
\begin{thebibliography}{36}%
\makeatletter
\providecommand \@ifxundefined [1]{%
 \@ifx{#1\undefined}
}%
\providecommand \@ifnum [1]{%
 \ifnum #1\expandafter \@firstoftwo
 \else \expandafter \@secondoftwo
 \fi
}%
\providecommand \@ifx [1]{%
 \ifx #1\expandafter \@firstoftwo
 \else \expandafter \@secondoftwo
 \fi
}%
\providecommand \natexlab [1]{#1}%
\providecommand \enquote  [1]{``#1''}%
\providecommand \bibnamefont  [1]{#1}%
\providecommand \bibfnamefont [1]{#1}%
\providecommand \citenamefont [1]{#1}%
\providecommand \href@noop [0]{\@secondoftwo}%
\providecommand \href [0]{\begingroup \@sanitize@url \@href}%
\providecommand \@href[1]{\@@startlink{#1}\@@href}%
\providecommand \@@href[1]{\endgroup#1\@@endlink}%
\providecommand \@sanitize@url [0]{\catcode `\\12\catcode `\$12\catcode
  `\&12\catcode `\#12\catcode `\^12\catcode `\_12\catcode `\%12\relax}%
\providecommand \@@startlink[1]{}%
\providecommand \@@endlink[0]{}%
\providecommand \url  [0]{\begingroup\@sanitize@url \@url }%
\providecommand \@url [1]{\endgroup\@href {#1}{\urlprefix }}%
\providecommand \urlprefix  [0]{URL }%
\providecommand \Eprint [0]{\href }%
\providecommand \doibase [0]{http://dx.doi.org/}%
\providecommand \selectlanguage [0]{\@gobble}%
\providecommand \bibinfo  [0]{\@secondoftwo}%
\providecommand \bibfield  [0]{\@secondoftwo}%
\providecommand \translation [1]{[#1]}%
\providecommand \BibitemOpen [0]{}%
\providecommand \bibitemStop [0]{}%
\providecommand \bibitemNoStop [0]{.\EOS\space}%
\providecommand \EOS [0]{\spacefactor3000\relax}%
\providecommand \BibitemShut  [1]{\csname bibitem#1\endcsname}%
\let\auto@bib@innerbib\@empty
\bibitem [{\citenamefont {Wang}\ \emph {et~al.}(2012)\citenamefont {Wang},
  \citenamefont {Shi}, \citenamefont {Hochberg}, \citenamefont {Adam},
  \citenamefont {Schelew}, \citenamefont {Young}, \citenamefont {Jaeger},\ and\
  \citenamefont {Chrostowski}}]{Litho1}%
  \BibitemOpen
  \bibfield  {author} {\bibinfo {author} {\bibfnamefont {X.}~\bibnamefont
  {Wang}}, \bibinfo {author} {\bibfnamefont {W.}~\bibnamefont {Shi}}, \bibinfo
  {author} {\bibfnamefont {M.}~\bibnamefont {Hochberg}}, \bibinfo {author}
  {\bibfnamefont {K.}~\bibnamefont {Adam}}, \bibinfo {author} {\bibfnamefont
  {E.}~\bibnamefont {Schelew}}, \bibinfo {author} {\bibfnamefont {J.~F.}\
  \bibnamefont {Young}}, \bibinfo {author} {\bibfnamefont {N.~A.~F.}\
  \bibnamefont {Jaeger}}, \ and\ \bibinfo {author} {\bibfnamefont
  {L.}~\bibnamefont {Chrostowski}},\ }\bibfield  {title} {\enquote {\bibinfo
  {title} {Lithography simulation for the fabrication of silicon photonic
  devices with deep-ultraviolet lithography},}\ }in\ \href {\doibase
  10.1109/GROUP4.2012.6324162} {\emph {\bibinfo {booktitle} {The 9th
  International Conference on Group IV Photonics (GFP)}}}\ (\bibinfo {year}
  {2012})\ pp.\ \bibinfo {pages} {288--290}\BibitemShut {NoStop}%
\bibitem [{\citenamefont {Lin}\ \emph {et~al.}(2020)\citenamefont {Lin},
  \citenamefont {Hammood}, \citenamefont {Yun}, \citenamefont {Luan},
  \citenamefont {Jaeger},\ and\ \citenamefont {Chrostowski}}]{Litho2}%
  \BibitemOpen
  \bibfield  {author} {\bibinfo {author} {\bibfnamefont {S.}~\bibnamefont
  {Lin}}, \bibinfo {author} {\bibfnamefont {M.}~\bibnamefont {Hammood}},
  \bibinfo {author} {\bibfnamefont {H.}~\bibnamefont {Yun}}, \bibinfo {author}
  {\bibfnamefont {E.}~\bibnamefont {Luan}}, \bibinfo {author} {\bibfnamefont
  {N.~A.~F.}\ \bibnamefont {Jaeger}}, \ and\ \bibinfo {author} {\bibfnamefont
  {L.}~\bibnamefont {Chrostowski}},\ }\bibfield  {title} {\enquote {\bibinfo
  {title} {Computational lithography for silicon photonics design},}\ }\href
  {\doibase 10.1109/JSTQE.2019.2958931} {\bibfield  {journal} {\bibinfo
  {journal} {IEEE Journal of Selected Topics in Quantum Electronics}\ }\textbf
  {\bibinfo {volume} {26}},\ \bibinfo {pages} {1--8} (\bibinfo {year}
  {2020})}\BibitemShut {NoStop}%
\bibitem [{\citenamefont {Yang}\ \emph {et~al.}(2022)\citenamefont {Yang},
  \citenamefont {Li}, \citenamefont {Sastry}, \citenamefont {Mukhopadhyay},
  \citenamefont {Kilgard}, \citenamefont {Anandkumar}, \citenamefont
  {Khailany}, \citenamefont {Singh},\ and\ \citenamefont {Ren}}]{NvidiaLitho}%
  \BibitemOpen
  \bibfield  {author} {\bibinfo {author} {\bibfnamefont {H.}~\bibnamefont
  {Yang}}, \bibinfo {author} {\bibfnamefont {Z.}~\bibnamefont {Li}}, \bibinfo
  {author} {\bibfnamefont {K.}~\bibnamefont {Sastry}}, \bibinfo {author}
  {\bibfnamefont {S.}~\bibnamefont {Mukhopadhyay}}, \bibinfo {author}
  {\bibfnamefont {M.}~\bibnamefont {Kilgard}}, \bibinfo {author} {\bibfnamefont
  {A.}~\bibnamefont {Anandkumar}}, \bibinfo {author} {\bibfnamefont
  {B.}~\bibnamefont {Khailany}}, \bibinfo {author} {\bibfnamefont
  {V.}~\bibnamefont {Singh}}, \ and\ \bibinfo {author} {\bibfnamefont
  {H.}~\bibnamefont {Ren}},\ }\href@noop {} {\enquote {\bibinfo {title}
  {Generic lithography modeling with dual-band optics-inspired neural
  networks},}\ } (\bibinfo {year} {2022}),\ \Eprint
  {http://arxiv.org/abs/2203.08616} {arXiv:2203.08616 [cs.OH]} \BibitemShut
  {NoStop}%
\bibitem [{\citenamefont {{Lecun}}\ \emph {et~al.}(1998)\citenamefont
  {{Lecun}}, \citenamefont {{Bottou}}, \citenamefont {{Bengio}},\ and\
  \citenamefont {{Haffner}}}]{originalimage}%
  \BibitemOpen
  \bibfield  {author} {\bibinfo {author} {\bibfnamefont {Y.}~\bibnamefont
  {{Lecun}}}, \bibinfo {author} {\bibfnamefont {L.}~\bibnamefont {{Bottou}}},
  \bibinfo {author} {\bibfnamefont {Y.}~\bibnamefont {{Bengio}}}, \ and\
  \bibinfo {author} {\bibfnamefont {P.}~\bibnamefont {{Haffner}}},\ }\bibfield
  {title} {\enquote {\bibinfo {title} {Gradient-based learning applied to
  document recognition},}\ }\href {\doibase 10.1109/5.726791} {\bibfield
  {journal} {\bibinfo  {journal} {Proceedings of the IEEE}\ }\textbf {\bibinfo
  {volume} {86}},\ \bibinfo {pages} {2278--2324} (\bibinfo {year}
  {1998})}\BibitemShut {NoStop}%
\bibitem [{\citenamefont {Ioffe}\ and\ \citenamefont
  {Szegedy}(2015)}]{Batchnormalization}%
  \BibitemOpen
  \bibfield  {author} {\bibinfo {author} {\bibfnamefont {S.}~\bibnamefont
  {Ioffe}}\ and\ \bibinfo {author} {\bibfnamefont {C.}~\bibnamefont
  {Szegedy}},\ }\bibfield  {title} {\enquote {\bibinfo {title} {Batch
  normalization: Accelerating deep network training by reducing internal
  covariate shift},}\ }\href {http://arxiv.org/abs/1502.03167} {\bibfield
  {journal} {\bibinfo  {journal} {CoRR}\ }\textbf {\bibinfo {volume}
  {abs/1502.03167}} (\bibinfo {year} {2015})},\ \Eprint
  {http://arxiv.org/abs/1502.03167} {arXiv:1502.03167} \BibitemShut {NoStop}%
\bibitem [{\citenamefont {He}\ \emph {et~al.}(2015)\citenamefont {He},
  \citenamefont {Zhang}, \citenamefont {Ren},\ and\ \citenamefont
  {Sun}}]{Resnet}%
  \BibitemOpen
  \bibfield  {author} {\bibinfo {author} {\bibfnamefont {K.}~\bibnamefont
  {He}}, \bibinfo {author} {\bibfnamefont {X.}~\bibnamefont {Zhang}}, \bibinfo
  {author} {\bibfnamefont {S.}~\bibnamefont {Ren}}, \ and\ \bibinfo {author}
  {\bibfnamefont {J.}~\bibnamefont {Sun}},\ }\bibfield  {title} {\enquote
  {\bibinfo {title} {Deep residual learning for image recognition},}\ }\href
  {http://arxiv.org/abs/1512.03385} {\bibfield  {journal} {\bibinfo  {journal}
  {CoRR}\ }\textbf {\bibinfo {volume} {abs/1512.03385}} (\bibinfo {year}
  {2015})},\ \Eprint {http://arxiv.org/abs/1512.03385} {arXiv:1512.03385}
  \BibitemShut {NoStop}%
\bibitem [{\citenamefont {Chalapathy}\ and\ \citenamefont
  {Chawla}(2019)}]{deepanomaly}%
  \BibitemOpen
  \bibfield  {author} {\bibinfo {author} {\bibfnamefont {R.}~\bibnamefont
  {Chalapathy}}\ and\ \bibinfo {author} {\bibfnamefont {S.}~\bibnamefont
  {Chawla}},\ }\bibfield  {title} {\enquote {\bibinfo {title} {Deep learning
  for anomaly detection: {A} survey},}\ }\href
  {http://arxiv.org/abs/1901.03407} {\bibfield  {journal} {\bibinfo  {journal}
  {CoRR}\ }\textbf {\bibinfo {volume} {abs/1901.03407}} (\bibinfo {year}
  {2019})},\ \Eprint {http://arxiv.org/abs/1901.03407} {arXiv:1901.03407}
  \BibitemShut {NoStop}%
\bibitem [{\citenamefont {Ouyang}\ \emph {et~al.}(2020)\citenamefont {Ouyang},
  \citenamefont {He}, \citenamefont {Ghorbani}, \citenamefont {Yuan},
  \citenamefont {Ebinger}, \citenamefont {Langlotz}, \citenamefont
  {Heidenreich}, \citenamefont {Harrington}, \citenamefont {Liang},
  \citenamefont {Ashley} \emph {et~al.}}]{cardiacml}%
  \BibitemOpen
  \bibfield  {author} {\bibinfo {author} {\bibfnamefont {D.}~\bibnamefont
  {Ouyang}}, \bibinfo {author} {\bibfnamefont {B.}~\bibnamefont {He}}, \bibinfo
  {author} {\bibfnamefont {A.}~\bibnamefont {Ghorbani}}, \bibinfo {author}
  {\bibfnamefont {N.}~\bibnamefont {Yuan}}, \bibinfo {author} {\bibfnamefont
  {J.}~\bibnamefont {Ebinger}}, \bibinfo {author} {\bibfnamefont {C.~P.}\
  \bibnamefont {Langlotz}}, \bibinfo {author} {\bibfnamefont {P.~A.}\
  \bibnamefont {Heidenreich}}, \bibinfo {author} {\bibfnamefont {R.~A.}\
  \bibnamefont {Harrington}}, \bibinfo {author} {\bibfnamefont {D.~H.}\
  \bibnamefont {Liang}}, \bibinfo {author} {\bibfnamefont {E.~A.}\ \bibnamefont
  {Ashley}},  \emph {et~al.},\ }\bibfield  {title} {\enquote {\bibinfo {title}
  {Video-based ai for beat-to-beat assessment of cardiac function},}\
  }\href@noop {} {\bibfield  {journal} {\bibinfo  {journal} {Nature}\ }\textbf
  {\bibinfo {volume} {580}},\ \bibinfo {pages} {252--256} (\bibinfo {year}
  {2020})}\BibitemShut {NoStop}%
\bibitem [{\citenamefont {Haenssle}\ \emph {et~al.}(2018)\citenamefont
  {Haenssle}, \citenamefont {Fink}, \citenamefont {Schneiderbauer},
  \citenamefont {Toberer}, \citenamefont {Buhl}, \citenamefont {Blum},
  \citenamefont {Kalloo}, \citenamefont {Hassen}, \citenamefont {Thomas},
  \citenamefont {Enk} \emph {et~al.}}]{medicalcomparison}%
  \BibitemOpen
  \bibfield  {author} {\bibinfo {author} {\bibfnamefont {H.~A.}\ \bibnamefont
  {Haenssle}}, \bibinfo {author} {\bibfnamefont {C.}~\bibnamefont {Fink}},
  \bibinfo {author} {\bibfnamefont {R.}~\bibnamefont {Schneiderbauer}},
  \bibinfo {author} {\bibfnamefont {F.}~\bibnamefont {Toberer}}, \bibinfo
  {author} {\bibfnamefont {T.}~\bibnamefont {Buhl}}, \bibinfo {author}
  {\bibfnamefont {A.}~\bibnamefont {Blum}}, \bibinfo {author} {\bibfnamefont
  {A.}~\bibnamefont {Kalloo}}, \bibinfo {author} {\bibfnamefont {A.~B.~H.}\
  \bibnamefont {Hassen}}, \bibinfo {author} {\bibfnamefont {L.}~\bibnamefont
  {Thomas}}, \bibinfo {author} {\bibfnamefont {A.}~\bibnamefont {Enk}},  \emph
  {et~al.},\ }\bibfield  {title} {\enquote {\bibinfo {title} {Man against
  machine: diagnostic performance of a deep learning convolutional neural
  network for dermoscopic melanoma recognition in comparison to 58
  dermatologists},}\ }\href@noop {} {\bibfield  {journal} {\bibinfo  {journal}
  {Annals of Oncology}\ }\textbf {\bibinfo {volume} {29}},\ \bibinfo {pages}
  {1836--1842} (\bibinfo {year} {2018})}\BibitemShut {NoStop}%
\bibitem [{\citenamefont {Tahersima}\ \emph {et~al.}(2019)\citenamefont
  {Tahersima}, \citenamefont {Kojima}, \citenamefont {Koike-Akino},
  \citenamefont {Jha}, \citenamefont {Wang}, \citenamefont {Lin},\ and\
  \citenamefont {Parsons}}]{deeplearningphotonics}%
  \BibitemOpen
  \bibfield  {author} {\bibinfo {author} {\bibfnamefont {M.~H.}\ \bibnamefont
  {Tahersima}}, \bibinfo {author} {\bibfnamefont {K.}~\bibnamefont {Kojima}},
  \bibinfo {author} {\bibfnamefont {T.}~\bibnamefont {Koike-Akino}}, \bibinfo
  {author} {\bibfnamefont {D.}~\bibnamefont {Jha}}, \bibinfo {author}
  {\bibfnamefont {B.}~\bibnamefont {Wang}}, \bibinfo {author} {\bibfnamefont
  {C.}~\bibnamefont {Lin}}, \ and\ \bibinfo {author} {\bibfnamefont
  {K.}~\bibnamefont {Parsons}},\ }\bibfield  {title} {\enquote {\bibinfo
  {title} {Deep neural network inverse design of integrated photonic power
  splitters},}\ }\href@noop {} {\bibfield  {journal} {\bibinfo  {journal}
  {Scientific reports}\ }\textbf {\bibinfo {volume} {9}},\ \bibinfo {pages}
  {1--9} (\bibinfo {year} {2019})}\BibitemShut {NoStop}%
\bibitem [{\citenamefont {Baker}\ \emph {et~al.}(2019)\citenamefont {Baker},
  \citenamefont {Kanitscheider}, \citenamefont {Markov}, \citenamefont {Wu},
  \citenamefont {Powell}, \citenamefont {McGrew},\ and\ \citenamefont
  {Mordatch}}]{openaihideandseek}%
  \BibitemOpen
  \bibfield  {author} {\bibinfo {author} {\bibfnamefont {B.}~\bibnamefont
  {Baker}}, \bibinfo {author} {\bibfnamefont {I.}~\bibnamefont
  {Kanitscheider}}, \bibinfo {author} {\bibfnamefont {T.~M.}\ \bibnamefont
  {Markov}}, \bibinfo {author} {\bibfnamefont {Y.}~\bibnamefont {Wu}}, \bibinfo
  {author} {\bibfnamefont {G.}~\bibnamefont {Powell}}, \bibinfo {author}
  {\bibfnamefont {B.}~\bibnamefont {McGrew}}, \ and\ \bibinfo {author}
  {\bibfnamefont {I.}~\bibnamefont {Mordatch}},\ }\bibfield  {title} {\enquote
  {\bibinfo {title} {Emergent tool use from multi-agent autocurricula},}\
  }\href {http://arxiv.org/abs/1909.07528} {\bibfield  {journal} {\bibinfo
  {journal} {CoRR}\ }\textbf {\bibinfo {volume} {abs/1909.07528}} (\bibinfo
  {year} {2019})},\ \Eprint {http://arxiv.org/abs/1909.07528}
  {arXiv:1909.07528} \BibitemShut {NoStop}%
\bibitem [{\citenamefont {Zheng}\ \emph {et~al.}(2020)\citenamefont {Zheng},
  \citenamefont {Trott}, \citenamefont {Srinivasa}, \citenamefont {Naik},
  \citenamefont {Gruesbeck}, \citenamefont {Parkes},\ and\ \citenamefont
  {Socher}}]{aieconomist}%
  \BibitemOpen
  \bibfield  {author} {\bibinfo {author} {\bibfnamefont {S.}~\bibnamefont
  {Zheng}}, \bibinfo {author} {\bibfnamefont {A.}~\bibnamefont {Trott}},
  \bibinfo {author} {\bibfnamefont {S.}~\bibnamefont {Srinivasa}}, \bibinfo
  {author} {\bibfnamefont {N.}~\bibnamefont {Naik}}, \bibinfo {author}
  {\bibfnamefont {M.}~\bibnamefont {Gruesbeck}}, \bibinfo {author}
  {\bibfnamefont {D.~C.}\ \bibnamefont {Parkes}}, \ and\ \bibinfo {author}
  {\bibfnamefont {R.}~\bibnamefont {Socher}},\ }\bibfield  {title} {\enquote
  {\bibinfo {title} {The ai economist: Improving equality and productivity with
  ai-driven tax policies},}\ }\href@noop {} {\bibfield  {journal} {\bibinfo
  {journal} {arXiv preprint arXiv:2004.13332}\ } (\bibinfo {year}
  {2020})}\BibitemShut {NoStop}%
\bibitem [{\citenamefont {OpenAI}\ \emph {et~al.}(2019)\citenamefont {OpenAI},
  \citenamefont {Akkaya}, \citenamefont {Andrychowicz}, \citenamefont
  {Chociej}, \citenamefont {Litwin}, \citenamefont {McGrew}, \citenamefont
  {Petron}, \citenamefont {Paino}, \citenamefont {Plappert}, \citenamefont
  {Powell}, \citenamefont {Ribas}, \citenamefont {Schneider}, \citenamefont
  {Tezak}, \citenamefont {Tworek}, \citenamefont {Welinder}, \citenamefont
  {Weng}, \citenamefont {Yuan}, \citenamefont {Zaremba},\ and\ \citenamefont
  {Zhang}}]{robotcube}%
  \BibitemOpen
  \bibfield  {author} {\bibinfo {author} {\bibnamefont {OpenAI}}, \bibinfo
  {author} {\bibfnamefont {I.}~\bibnamefont {Akkaya}}, \bibinfo {author}
  {\bibfnamefont {M.}~\bibnamefont {Andrychowicz}}, \bibinfo {author}
  {\bibfnamefont {M.}~\bibnamefont {Chociej}}, \bibinfo {author} {\bibfnamefont
  {M.}~\bibnamefont {Litwin}}, \bibinfo {author} {\bibfnamefont
  {B.}~\bibnamefont {McGrew}}, \bibinfo {author} {\bibfnamefont
  {A.}~\bibnamefont {Petron}}, \bibinfo {author} {\bibfnamefont
  {A.}~\bibnamefont {Paino}}, \bibinfo {author} {\bibfnamefont
  {M.}~\bibnamefont {Plappert}}, \bibinfo {author} {\bibfnamefont
  {G.}~\bibnamefont {Powell}}, \bibinfo {author} {\bibfnamefont
  {R.}~\bibnamefont {Ribas}}, \bibinfo {author} {\bibfnamefont
  {J.}~\bibnamefont {Schneider}}, \bibinfo {author} {\bibfnamefont
  {N.}~\bibnamefont {Tezak}}, \bibinfo {author} {\bibfnamefont
  {J.}~\bibnamefont {Tworek}}, \bibinfo {author} {\bibfnamefont
  {P.}~\bibnamefont {Welinder}}, \bibinfo {author} {\bibfnamefont
  {L.}~\bibnamefont {Weng}}, \bibinfo {author} {\bibfnamefont {Q.}~\bibnamefont
  {Yuan}}, \bibinfo {author} {\bibfnamefont {W.}~\bibnamefont {Zaremba}}, \
  and\ \bibinfo {author} {\bibfnamefont {L.}~\bibnamefont {Zhang}},\ }\bibfield
   {title} {\enquote {\bibinfo {title} {Solving rubik's cube with a robot
  hand},}\ }\href {http://arxiv.org/abs/1910.07113} {\bibfield  {journal}
  {\bibinfo  {journal} {CoRR}\ }\textbf {\bibinfo {volume} {abs/1910.07113}}
  (\bibinfo {year} {2019})},\ \Eprint {http://arxiv.org/abs/1910.07113}
  {arXiv:1910.07113} \BibitemShut {NoStop}%
\bibitem [{\citenamefont {Lee}\ \emph {et~al.}(2020)\citenamefont {Lee},
  \citenamefont {Hwangbo}, \citenamefont {Wellhausen}, \citenamefont {Koltun},\
  and\ \citenamefont {Hutter}}]{robotwalking}%
  \BibitemOpen
  \bibfield  {author} {\bibinfo {author} {\bibfnamefont {J.}~\bibnamefont
  {Lee}}, \bibinfo {author} {\bibfnamefont {J.}~\bibnamefont {Hwangbo}},
  \bibinfo {author} {\bibfnamefont {L.}~\bibnamefont {Wellhausen}}, \bibinfo
  {author} {\bibfnamefont {V.}~\bibnamefont {Koltun}}, \ and\ \bibinfo {author}
  {\bibfnamefont {M.}~\bibnamefont {Hutter}},\ }\bibfield  {title} {\enquote
  {\bibinfo {title} {Learning quadrupedal locomotion over challenging
  terrain},}\ }\href {\doibase 10.1126/scirobotics.abc5986} {\bibfield
  {journal} {\bibinfo  {journal} {Science Robotics}\ }\textbf {\bibinfo
  {volume} {5}},\ \bibinfo {pages} {eabc5986} (\bibinfo {year}
  {2020})}\BibitemShut {NoStop}%
\bibitem [{\citenamefont {Sui}\ \emph {et~al.}(2021)\citenamefont {Sui},
  \citenamefont {Guo}, \citenamefont {Zhang}, \citenamefont {Gu},\ and\
  \citenamefont {Lin}}]{RLmaterials}%
  \BibitemOpen
  \bibfield  {author} {\bibinfo {author} {\bibfnamefont {F.}~\bibnamefont
  {Sui}}, \bibinfo {author} {\bibfnamefont {R.}~\bibnamefont {Guo}}, \bibinfo
  {author} {\bibfnamefont {Z.}~\bibnamefont {Zhang}}, \bibinfo {author}
  {\bibfnamefont {G.~X.}\ \bibnamefont {Gu}}, \ and\ \bibinfo {author}
  {\bibfnamefont {L.}~\bibnamefont {Lin}},\ }\bibfield  {title} {\enquote
  {\bibinfo {title} {Deep reinforcement learning for digital materials
  design},}\ }\href {\doibase 10.1021/acsmaterialslett.1c00390} {\bibfield
  {journal} {\bibinfo  {journal} {ACS Materials Letters}\ }\textbf {\bibinfo
  {volume} {3}},\ \bibinfo {pages} {1433--1439} (\bibinfo {year} {2021})},\
  \Eprint
  {http://arxiv.org/abs/https://doi.org/10.1021/acsmaterialslett.1c00390}
  {https://doi.org/10.1021/acsmaterialslett.1c00390} \BibitemShut {NoStop}%
\bibitem [{\citenamefont {So}\ \emph {et~al.}(2020)\citenamefont {So},
  \citenamefont {Badloe}, \citenamefont {Noh}, \citenamefont {Bravo-Abad},\
  and\ \citenamefont {Rho}}]{RLphotonics}%
  \BibitemOpen
  \bibfield  {author} {\bibinfo {author} {\bibfnamefont {S.}~\bibnamefont
  {So}}, \bibinfo {author} {\bibfnamefont {T.}~\bibnamefont {Badloe}}, \bibinfo
  {author} {\bibfnamefont {J.}~\bibnamefont {Noh}}, \bibinfo {author}
  {\bibfnamefont {J.}~\bibnamefont {Bravo-Abad}}, \ and\ \bibinfo {author}
  {\bibfnamefont {J.}~\bibnamefont {Rho}},\ }\bibfield  {title} {\enquote
  {\bibinfo {title} {Deep learning enabled inverse design in nanophotonics},}\
  }\href {\doibase doi:10.1515/nanoph-2019-0474} {\bibfield  {journal}
  {\bibinfo  {journal} {Nanophotonics}\ }\textbf {\bibinfo {volume} {9}},\
  \bibinfo {pages} {1041--1057} (\bibinfo {year} {2020})}\BibitemShut {NoStop}%
\bibitem [{\citenamefont {Li}\ \emph {et~al.}(2023)\citenamefont {Li},
  \citenamefont {Zhang}, \citenamefont {Xie}, \citenamefont {Gong},
  \citenamefont {Ding}, \citenamefont {Dai}, \citenamefont {Chen},
  \citenamefont {Yin},\ and\ \citenamefont {Zhang}}]{RLphotonics2}%
  \BibitemOpen
  \bibfield  {author} {\bibinfo {author} {\bibfnamefont {R.}~\bibnamefont
  {Li}}, \bibinfo {author} {\bibfnamefont {C.}~\bibnamefont {Zhang}}, \bibinfo
  {author} {\bibfnamefont {W.}~\bibnamefont {Xie}}, \bibinfo {author}
  {\bibfnamefont {Y.}~\bibnamefont {Gong}}, \bibinfo {author} {\bibfnamefont
  {F.}~\bibnamefont {Ding}}, \bibinfo {author} {\bibfnamefont {H.}~\bibnamefont
  {Dai}}, \bibinfo {author} {\bibfnamefont {Z.}~\bibnamefont {Chen}}, \bibinfo
  {author} {\bibfnamefont {F.}~\bibnamefont {Yin}}, \ and\ \bibinfo {author}
  {\bibfnamefont {Z.}~\bibnamefont {Zhang}},\ }\bibfield  {title} {\enquote
  {\bibinfo {title} {Deep reinforcement learning empowers automated inverse
  design and optimization of photonic crystals for nanoscale laser cavities},}\
  }\href {\doibase doi:10.1515/nanoph-2022-0692} {\bibfield  {journal}
  {\bibinfo  {journal} {Nanophotonics}\ }\textbf {\bibinfo {volume} {12}},\
  \bibinfo {pages} {319--334} (\bibinfo {year} {2023})}\BibitemShut {NoStop}%
\bibitem [{\citenamefont {Roy}\ \emph {et~al.}(2021)\citenamefont {Roy},
  \citenamefont {Raiman}, \citenamefont {Kant}, \citenamefont {Elkin},
  \citenamefont {Kirby}, \citenamefont {Siu}, \citenamefont {Oberman},
  \citenamefont {Godil},\ and\ \citenamefont {Catanzaro}}]{Nvidia}%
  \BibitemOpen
  \bibfield  {author} {\bibinfo {author} {\bibfnamefont {R.}~\bibnamefont
  {Roy}}, \bibinfo {author} {\bibfnamefont {J.}~\bibnamefont {Raiman}},
  \bibinfo {author} {\bibfnamefont {N.}~\bibnamefont {Kant}}, \bibinfo {author}
  {\bibfnamefont {I.}~\bibnamefont {Elkin}}, \bibinfo {author} {\bibfnamefont
  {R.}~\bibnamefont {Kirby}}, \bibinfo {author} {\bibfnamefont
  {M.}~\bibnamefont {Siu}}, \bibinfo {author} {\bibfnamefont {S.}~\bibnamefont
  {Oberman}}, \bibinfo {author} {\bibfnamefont {S.}~\bibnamefont {Godil}}, \
  and\ \bibinfo {author} {\bibfnamefont {B.}~\bibnamefont {Catanzaro}},\
  }\bibfield  {title} {\enquote {\bibinfo {title} {Prefixrl: Optimization of
  parallel prefix circuits using deep reinforcement learning},}\ }in\ \href
  {\doibase 10.1109/DAC18074.2021.9586094} {\emph {\bibinfo {booktitle} {2021
  58th ACM/IEEE Design Automation Conference (DAC)}}}\ (\bibinfo {year}
  {2021})\ pp.\ \bibinfo {pages} {853--858}\BibitemShut {NoStop}%
\bibitem [{\citenamefont {Yan}\ \emph {et~al.}(2021)\citenamefont {Yan},
  \citenamefont {Gitt}, \citenamefont {Lin}, \citenamefont {Witt},
  \citenamefont {Abdolahi}, \citenamefont {Afifi}, \citenamefont {Azem},
  \citenamefont {Darcie}, \citenamefont {Wu}, \citenamefont {Awan},
  \citenamefont {Mitchell}, \citenamefont {Pfenning}, \citenamefont
  {Chrostowski},\ and\ \citenamefont {Young}}]{quantumReview}%
  \BibitemOpen
  \bibfield  {author} {\bibinfo {author} {\bibfnamefont {X.}~\bibnamefont
  {Yan}}, \bibinfo {author} {\bibfnamefont {S.}~\bibnamefont {Gitt}}, \bibinfo
  {author} {\bibfnamefont {B.}~\bibnamefont {Lin}}, \bibinfo {author}
  {\bibfnamefont {D.}~\bibnamefont {Witt}}, \bibinfo {author} {\bibfnamefont
  {M.}~\bibnamefont {Abdolahi}}, \bibinfo {author} {\bibfnamefont
  {A.}~\bibnamefont {Afifi}}, \bibinfo {author} {\bibfnamefont
  {A.}~\bibnamefont {Azem}}, \bibinfo {author} {\bibfnamefont {A.}~\bibnamefont
  {Darcie}}, \bibinfo {author} {\bibfnamefont {J.}~\bibnamefont {Wu}}, \bibinfo
  {author} {\bibfnamefont {K.}~\bibnamefont {Awan}}, \bibinfo {author}
  {\bibfnamefont {M.}~\bibnamefont {Mitchell}}, \bibinfo {author}
  {\bibfnamefont {A.}~\bibnamefont {Pfenning}}, \bibinfo {author}
  {\bibfnamefont {L.}~\bibnamefont {Chrostowski}}, \ and\ \bibinfo {author}
  {\bibfnamefont {J.~F.}\ \bibnamefont {Young}},\ }\bibfield  {title} {\enquote
  {\bibinfo {title} {Silicon photonic quantum computing with spin qubits},}\
  }\href {\doibase 10.1063/5.0049372} {\bibfield  {journal} {\bibinfo
  {journal} {APL Photonics}\ }\textbf {\bibinfo {volume} {6}},\ \bibinfo
  {pages} {070901} (\bibinfo {year} {2021})},\ \Eprint
  {http://arxiv.org/abs/https://doi.org/10.1063/5.0049372}
  {https://doi.org/10.1063/5.0049372} \BibitemShut {NoStop}%
\bibitem [{\citenamefont {Dhaliah}\ \emph {et~al.}(2022)\citenamefont
  {Dhaliah}, \citenamefont {Xiong}, \citenamefont {Sipahigil}, \citenamefont
  {Griffin},\ and\ \citenamefont {Hautier}}]{Tcenter}%
  \BibitemOpen
  \bibfield  {author} {\bibinfo {author} {\bibfnamefont {D.}~\bibnamefont
  {Dhaliah}}, \bibinfo {author} {\bibfnamefont {Y.}~\bibnamefont {Xiong}},
  \bibinfo {author} {\bibfnamefont {A.}~\bibnamefont {Sipahigil}}, \bibinfo
  {author} {\bibfnamefont {S.~M.}\ \bibnamefont {Griffin}}, \ and\ \bibinfo
  {author} {\bibfnamefont {G.}~\bibnamefont {Hautier}},\ }\bibfield  {title}
  {\enquote {\bibinfo {title} {First-principles study of the t center in
  silicon},}\ }\href {\doibase 10.1103/PhysRevMaterials.6.L053201} {\bibfield
  {journal} {\bibinfo  {journal} {Phys. Rev. Materials}\ }\textbf {\bibinfo
  {volume} {6}},\ \bibinfo {pages} {L053201} (\bibinfo {year}
  {2022})}\BibitemShut {NoStop}%
\bibitem [{\citenamefont {Baron}\ \emph {et~al.}(2022)\citenamefont {Baron},
  \citenamefont {Durand}, \citenamefont {Udvarhelyi}, \citenamefont {Herzig},
  \citenamefont {Khoury}, \citenamefont {Pezzagna}, \citenamefont {Meijer},
  \citenamefont {Robert-Philip}, \citenamefont {Abbarchi}, \citenamefont
  {Hartmann}, \citenamefont {Mazzocchi}, \citenamefont {Gérard}, \citenamefont
  {Gali}, \citenamefont {Jacques}, \citenamefont {Cassabois},\ and\
  \citenamefont {Dréau}}]{Wcenter}%
  \BibitemOpen
  \bibfield  {author} {\bibinfo {author} {\bibfnamefont {Y.}~\bibnamefont
  {Baron}}, \bibinfo {author} {\bibfnamefont {A.}~\bibnamefont {Durand}},
  \bibinfo {author} {\bibfnamefont {P.}~\bibnamefont {Udvarhelyi}}, \bibinfo
  {author} {\bibfnamefont {T.}~\bibnamefont {Herzig}}, \bibinfo {author}
  {\bibfnamefont {M.}~\bibnamefont {Khoury}}, \bibinfo {author} {\bibfnamefont
  {S.}~\bibnamefont {Pezzagna}}, \bibinfo {author} {\bibfnamefont
  {J.}~\bibnamefont {Meijer}}, \bibinfo {author} {\bibfnamefont
  {I.}~\bibnamefont {Robert-Philip}}, \bibinfo {author} {\bibfnamefont
  {M.}~\bibnamefont {Abbarchi}}, \bibinfo {author} {\bibfnamefont {J.-M.}\
  \bibnamefont {Hartmann}}, \bibinfo {author} {\bibfnamefont {V.}~\bibnamefont
  {Mazzocchi}}, \bibinfo {author} {\bibfnamefont {J.-M.}\ \bibnamefont
  {Gérard}}, \bibinfo {author} {\bibfnamefont {A.}~\bibnamefont {Gali}},
  \bibinfo {author} {\bibfnamefont {V.}~\bibnamefont {Jacques}}, \bibinfo
  {author} {\bibfnamefont {G.}~\bibnamefont {Cassabois}}, \ and\ \bibinfo
  {author} {\bibfnamefont {A.}~\bibnamefont {Dréau}},\ }\bibfield  {title}
  {\enquote {\bibinfo {title} {Detection of single w-centers in silicon},}\
  }\href {\doibase 10.1021/acsphotonics.2c00336} {\bibfield  {journal}
  {\bibinfo  {journal} {ACS Photonics}\ }\textbf {\bibinfo {volume} {9}},\
  \bibinfo {pages} {2337--2345} (\bibinfo {year} {2022})},\ \Eprint
  {http://arxiv.org/abs/https://doi.org/10.1021/acsphotonics.2c00336}
  {https://doi.org/10.1021/acsphotonics.2c00336} \BibitemShut {NoStop}%
\bibitem [{\citenamefont {Darcie}\ \emph {et~al.}(2021)\citenamefont {Darcie},
  \citenamefont {Mitchell}, \citenamefont {Awan}, \citenamefont {Abdolahi},
  \citenamefont {Hammood}, \citenamefont {Pfenning}, \citenamefont {Yan},
  \citenamefont {Afifi}, \citenamefont {Witt}, \citenamefont {Lin},
  \citenamefont {Gou}, \citenamefont {Jhoja}, \citenamefont {Wu}, \citenamefont
  {Taghavi}, \citenamefont {Weekes}, \citenamefont {Jaeger}, \citenamefont
  {Young},\ and\ \citenamefont {Chrostowski}}]{SiEPICFab}%
  \BibitemOpen
  \bibfield  {author} {\bibinfo {author} {\bibfnamefont {A.}~\bibnamefont
  {Darcie}}, \bibinfo {author} {\bibfnamefont {M.}~\bibnamefont {Mitchell}},
  \bibinfo {author} {\bibfnamefont {K.~M.}\ \bibnamefont {Awan}}, \bibinfo
  {author} {\bibfnamefont {M.}~\bibnamefont {Abdolahi}}, \bibinfo {author}
  {\bibfnamefont {M.}~\bibnamefont {Hammood}}, \bibinfo {author} {\bibfnamefont
  {A.}~\bibnamefont {Pfenning}}, \bibinfo {author} {\bibfnamefont
  {X.}~\bibnamefont {Yan}}, \bibinfo {author} {\bibfnamefont {A.~E.}\
  \bibnamefont {Afifi}}, \bibinfo {author} {\bibfnamefont {D.~M.}\ \bibnamefont
  {Witt}}, \bibinfo {author} {\bibfnamefont {B.}~\bibnamefont {Lin}}, \bibinfo
  {author} {\bibfnamefont {S.}~\bibnamefont {Gou}}, \bibinfo {author}
  {\bibfnamefont {J.}~\bibnamefont {Jhoja}}, \bibinfo {author} {\bibfnamefont
  {J.}~\bibnamefont {Wu}}, \bibinfo {author} {\bibfnamefont {I.}~\bibnamefont
  {Taghavi}}, \bibinfo {author} {\bibfnamefont {D.~M.}\ \bibnamefont {Weekes}},
  \bibinfo {author} {\bibfnamefont {N.~A.~F.}\ \bibnamefont {Jaeger}}, \bibinfo
  {author} {\bibfnamefont {J.~F.}\ \bibnamefont {Young}}, \ and\ \bibinfo
  {author} {\bibfnamefont {L.}~\bibnamefont {Chrostowski}},\ }\bibfield
  {title} {\enquote {\bibinfo {title} {Siepicfab: the canadian silicon
  photonics rapid-prototyping foundry for integrated optics and quantum
  computing},}\ }in\ \href {\doibase 10.1117/12.2583432} {\emph {\bibinfo
  {booktitle} {Silicon Photonics XVI}}},\ Vol.\ \bibinfo {volume} {11691}\
  (\bibinfo  {publisher} {SPIE},\ \bibinfo {year} {2021})\ pp.\ \bibinfo
  {pages} {31--50},\ \bibinfo {note}
  {\url{https://www.spiedigitallibrary.org/conference-proceedings-of-spie/11691/116910C/SiEPICfab--the-Canadian-silicon-photonics-rapid-prototyping-foundry-for/10.1117/12.2583432.full}}\BibitemShut
  {NoStop}%
\bibitem [{\citenamefont {Halir}\ \emph {et~al.}(2009)\citenamefont {Halir},
  \citenamefont {Cheben}, \citenamefont {Janz}, \citenamefont {Xu},
  \citenamefont {{n}igo Molina-Fern\'{a}ndez},\ and\ \citenamefont
  {Wang\"{u}emert-P\'{e}rez}}]{NRC_gc}%
  \BibitemOpen
  \bibfield  {author} {\bibinfo {author} {\bibfnamefont {R.}~\bibnamefont
  {Halir}}, \bibinfo {author} {\bibfnamefont {P.}~\bibnamefont {Cheben}},
  \bibinfo {author} {\bibfnamefont {S.}~\bibnamefont {Janz}}, \bibinfo {author}
  {\bibfnamefont {D.-X.}\ \bibnamefont {Xu}}, \bibinfo {author} {\bibfnamefont
  {I.}~\bibnamefont {{n}igo Molina-Fern\'{a}ndez}}, \ and\ \bibinfo {author}
  {\bibfnamefont {J.~G.}\ \bibnamefont {Wang\"{u}emert-P\'{e}rez}},\ }\bibfield
   {title} {\enquote {\bibinfo {title} {Waveguide grating coupler with
  subwavelength microstructures},}\ }\href {\doibase 10.1364/OL.34.001408}
  {\bibfield  {journal} {\bibinfo  {journal} {Opt. Lett.}\ }\textbf {\bibinfo
  {volume} {34}},\ \bibinfo {pages} {1408--1410} (\bibinfo {year}
  {2009})}\BibitemShut {NoStop}%
\bibitem [{\citenamefont {Wang}\ \emph {et~al.}(2014)\citenamefont {Wang},
  \citenamefont {Wang}, \citenamefont {Flueckiger}, \citenamefont {Yun},
  \citenamefont {Shi}, \citenamefont {Bojko}, \citenamefont {Jaeger},\ and\
  \citenamefont {Chrostowski}}]{subwave}%
  \BibitemOpen
  \bibfield  {author} {\bibinfo {author} {\bibfnamefont {Y.}~\bibnamefont
  {Wang}}, \bibinfo {author} {\bibfnamefont {X.}~\bibnamefont {Wang}}, \bibinfo
  {author} {\bibfnamefont {J.}~\bibnamefont {Flueckiger}}, \bibinfo {author}
  {\bibfnamefont {H.}~\bibnamefont {Yun}}, \bibinfo {author} {\bibfnamefont
  {W.}~\bibnamefont {Shi}}, \bibinfo {author} {\bibfnamefont {R.}~\bibnamefont
  {Bojko}}, \bibinfo {author} {\bibfnamefont {N.~A.~F.}\ \bibnamefont
  {Jaeger}}, \ and\ \bibinfo {author} {\bibfnamefont {L.}~\bibnamefont
  {Chrostowski}},\ }\bibfield  {title} {\enquote {\bibinfo {title} {Focusing
  sub-wavelength grating couplers with low back reflections for rapid
  prototyping of silicon photonic circuits},}\ }\href {\doibase
  10.1364/OE.22.020652} {\bibfield  {journal} {\bibinfo  {journal} {Opt.
  Express}\ }\textbf {\bibinfo {volume} {22}},\ \bibinfo {pages} {20652--20662}
  (\bibinfo {year} {2014})}\BibitemShut {NoStop}%
\bibitem [{\citenamefont {Liu}\ \emph {et~al.}(2010)\citenamefont {Liu},
  \citenamefont {Pu}, \citenamefont {Yvind},\ and\ \citenamefont
  {Hvam}}]{normphcgc}%
  \BibitemOpen
  \bibfield  {author} {\bibinfo {author} {\bibfnamefont {L.}~\bibnamefont
  {Liu}}, \bibinfo {author} {\bibfnamefont {M.}~\bibnamefont {Pu}}, \bibinfo
  {author} {\bibfnamefont {K.}~\bibnamefont {Yvind}}, \ and\ \bibinfo {author}
  {\bibfnamefont {J.~M.}\ \bibnamefont {Hvam}},\ }\bibfield  {title} {\enquote
  {\bibinfo {title} {High-efficiency, large-bandwidth silicon-on-insulator
  grating coupler based on a fully-etched photonic crystal structure},}\ }\href
  {\doibase 10.1063/1.3304791} {\bibfield  {journal} {\bibinfo  {journal}
  {Applied Physics Letters}\ }\textbf {\bibinfo {volume} {96}},\ \bibinfo
  {pages} {051126} (\bibinfo {year} {2010})},\ \Eprint
  {http://arxiv.org/abs/https://doi.org/10.1063/1.3304791}
  {https://doi.org/10.1063/1.3304791} \BibitemShut {NoStop}%
\bibitem [{\citenamefont {Ding}, \citenamefont {Ou},\ and\ \citenamefont
  {Peucheret}(2013)}]{phcgcap}%
  \BibitemOpen
  \bibfield  {author} {\bibinfo {author} {\bibfnamefont {Y.}~\bibnamefont
  {Ding}}, \bibinfo {author} {\bibfnamefont {H.}~\bibnamefont {Ou}}, \ and\
  \bibinfo {author} {\bibfnamefont {C.}~\bibnamefont {Peucheret}},\ }\bibfield
  {title} {\enquote {\bibinfo {title} {Ultrahigh-efficiency apodized grating
  coupler using fully etched photonic crystals},}\ }\href {\doibase
  10.1364/OL.38.002732} {\bibfield  {journal} {\bibinfo  {journal} {Opt.
  Lett.}\ }\textbf {\bibinfo {volume} {38}},\ \bibinfo {pages} {2732--2734}
  (\bibinfo {year} {2013})}\BibitemShut {NoStop}%
\bibitem [{\citenamefont {Lillicrap}\ \emph {et~al.}(2019)\citenamefont
  {Lillicrap}, \citenamefont {Hunt}, \citenamefont {Pritzel}, \citenamefont
  {Heess}, \citenamefont {Erez}, \citenamefont {Tassa}, \citenamefont
  {Silver},\ and\ \citenamefont {Wierstra}}]{ddpgpaper}%
  \BibitemOpen
  \bibfield  {author} {\bibinfo {author} {\bibfnamefont {T.~P.}\ \bibnamefont
  {Lillicrap}}, \bibinfo {author} {\bibfnamefont {J.~J.}\ \bibnamefont {Hunt}},
  \bibinfo {author} {\bibfnamefont {A.}~\bibnamefont {Pritzel}}, \bibinfo
  {author} {\bibfnamefont {N.}~\bibnamefont {Heess}}, \bibinfo {author}
  {\bibfnamefont {T.}~\bibnamefont {Erez}}, \bibinfo {author} {\bibfnamefont
  {Y.}~\bibnamefont {Tassa}}, \bibinfo {author} {\bibfnamefont
  {D.}~\bibnamefont {Silver}}, \ and\ \bibinfo {author} {\bibfnamefont
  {D.}~\bibnamefont {Wierstra}},\ }\href@noop {} {\enquote {\bibinfo {title}
  {Continuous control with deep reinforcement learning},}\ } (\bibinfo {year}
  {2019}),\ \Eprint {http://arxiv.org/abs/1509.02971} {arXiv:1509.02971
  [cs.LG]} \BibitemShut {NoStop}%
\bibitem [{\citenamefont {Tabor}(2020)}]{ddpgcode}%
  \BibitemOpen
  \bibfield  {author} {\bibinfo {author} {\bibfnamefont {P.}~\bibnamefont
  {Tabor}},\ }\href@noop {} {\enquote {\bibinfo {title}
  {Actor-critic-methods-paper-to-code},}\ }\bibinfo {howpublished}
  {\url{https://github.com/philtabor/Actor-Critic-Methods-Paper-To-Code/tree/master/DDPG}}
  (\bibinfo {year} {2020}),\ \bibinfo {note} {[retrieved 5 June
  2020]}\BibitemShut {NoStop}%
\bibitem [{\citenamefont {Rumelhart}, \citenamefont {Hinton},\ and\
  \citenamefont {Williams}(1986)}]{backprop}%
  \BibitemOpen
  \bibfield  {author} {\bibinfo {author} {\bibfnamefont {D.~E.}\ \bibnamefont
  {Rumelhart}}, \bibinfo {author} {\bibfnamefont {G.~E.}\ \bibnamefont
  {Hinton}}, \ and\ \bibinfo {author} {\bibfnamefont {R.~J.}\ \bibnamefont
  {Williams}},\ }\bibfield  {title} {\enquote {\bibinfo {title} {Learning
  representations by back-propagating errors},}\ }\href@noop {} {\bibfield
  {journal} {\bibinfo  {journal} {Nature}\ }\textbf {\bibinfo {volume} {323}},\
  \bibinfo {pages} {533--536} (\bibinfo {year} {1986})}\BibitemShut {NoStop}%
\bibitem [{\citenamefont {Shi}\ \emph {et~al.}(2013)\citenamefont {Shi},
  \citenamefont {Yun}, \citenamefont {Lin}, \citenamefont {Greenberg},
  \citenamefont {Wang}, \citenamefont {Wang}, \citenamefont {Fard},
  \citenamefont {Flueckiger}, \citenamefont {Jaeger},\ and\ \citenamefont
  {Chrostowski}}]{contra-directional}%
  \BibitemOpen
  \bibfield  {author} {\bibinfo {author} {\bibfnamefont {W.}~\bibnamefont
  {Shi}}, \bibinfo {author} {\bibfnamefont {H.}~\bibnamefont {Yun}}, \bibinfo
  {author} {\bibfnamefont {C.}~\bibnamefont {Lin}}, \bibinfo {author}
  {\bibfnamefont {M.}~\bibnamefont {Greenberg}}, \bibinfo {author}
  {\bibfnamefont {X.}~\bibnamefont {Wang}}, \bibinfo {author} {\bibfnamefont
  {Y.}~\bibnamefont {Wang}}, \bibinfo {author} {\bibfnamefont {S.~T.}\
  \bibnamefont {Fard}}, \bibinfo {author} {\bibfnamefont {J.}~\bibnamefont
  {Flueckiger}}, \bibinfo {author} {\bibfnamefont {N.~A.~F.}\ \bibnamefont
  {Jaeger}}, \ and\ \bibinfo {author} {\bibfnamefont {L.}~\bibnamefont
  {Chrostowski}},\ }\bibfield  {title} {\enquote {\bibinfo {title}
  {Ultra-compact, flat-top demultiplexer using anti-reflection
  contra-directional couplers for cwdm networks on silicon},}\ }\href {\doibase
  10.1364/OE.21.006733} {\bibfield  {journal} {\bibinfo  {journal} {Opt.
  Express}\ }\textbf {\bibinfo {volume} {21}},\ \bibinfo {pages} {6733--6738}
  (\bibinfo {year} {2013})}\BibitemShut {NoStop}%
\bibitem [{\citenamefont {Zhang}\ \emph {et~al.}(2013)\citenamefont {Zhang},
  \citenamefont {Yang}, \citenamefont {Lim}, \citenamefont {Lo}, \citenamefont
  {Galland}, \citenamefont {Baehr-Jones},\ and\ \citenamefont
  {Hochberg}}]{ybranch}%
  \BibitemOpen
  \bibfield  {author} {\bibinfo {author} {\bibfnamefont {Y.}~\bibnamefont
  {Zhang}}, \bibinfo {author} {\bibfnamefont {S.}~\bibnamefont {Yang}},
  \bibinfo {author} {\bibfnamefont {A.~E.-J.}\ \bibnamefont {Lim}}, \bibinfo
  {author} {\bibfnamefont {G.-Q.}\ \bibnamefont {Lo}}, \bibinfo {author}
  {\bibfnamefont {C.}~\bibnamefont {Galland}}, \bibinfo {author} {\bibfnamefont
  {T.}~\bibnamefont {Baehr-Jones}}, \ and\ \bibinfo {author} {\bibfnamefont
  {M.}~\bibnamefont {Hochberg}},\ }\bibfield  {title} {\enquote {\bibinfo
  {title} {A compact and low loss y-junction for submicron silicon
  waveguide},}\ }\href {\doibase 10.1364/OE.21.001310} {\bibfield  {journal}
  {\bibinfo  {journal} {Opt. Express}\ }\textbf {\bibinfo {volume} {21}},\
  \bibinfo {pages} {1310--1316} (\bibinfo {year} {2013})}\BibitemShut {NoStop}%
\bibitem [{\citenamefont {Shankar}\ \emph {et~al.}(2011)\citenamefont
  {Shankar}, \citenamefont {Leijssen}, \citenamefont {Bulu},\ and\
  \citenamefont {Lon\v{c}ar}}]{mircavity}%
  \BibitemOpen
  \bibfield  {author} {\bibinfo {author} {\bibfnamefont {R.}~\bibnamefont
  {Shankar}}, \bibinfo {author} {\bibfnamefont {R.}~\bibnamefont {Leijssen}},
  \bibinfo {author} {\bibfnamefont {I.}~\bibnamefont {Bulu}}, \ and\ \bibinfo
  {author} {\bibfnamefont {M.}~\bibnamefont {Lon\v{c}ar}},\ }\bibfield  {title}
  {\enquote {\bibinfo {title} {Mid-infrared photonic crystal cavities in
  silicon},}\ }\href {\doibase 10.1364/OE.19.005579} {\bibfield  {journal}
  {\bibinfo  {journal} {Opt. Express}\ }\textbf {\bibinfo {volume} {19}},\
  \bibinfo {pages} {5579--5586} (\bibinfo {year} {2011})}\BibitemShut {NoStop}%
\bibitem [{\citenamefont {Choi}, \citenamefont {Heuck},\ and\ \citenamefont
  {Englund}(2017)}]{nanobeam}%
  \BibitemOpen
  \bibfield  {author} {\bibinfo {author} {\bibfnamefont {H.}~\bibnamefont
  {Choi}}, \bibinfo {author} {\bibfnamefont {M.}~\bibnamefont {Heuck}}, \ and\
  \bibinfo {author} {\bibfnamefont {D.}~\bibnamefont {Englund}},\ }\bibfield
  {title} {\enquote {\bibinfo {title} {Self-similar nanocavity design with
  ultrasmall mode volume for single-photon nonlinearities},}\ }\href {\doibase
  10.1103/PhysRevLett.118.223605} {\bibfield  {journal} {\bibinfo  {journal}
  {Phys. Rev. Lett.}\ }\textbf {\bibinfo {volume} {118}},\ \bibinfo {pages}
  {223605} (\bibinfo {year} {2017})}\BibitemShut {NoStop}%
\bibitem [{\citenamefont {Matsuo}\ \emph {et~al.}(2012)\citenamefont {Matsuo},
  \citenamefont {Takeda}, \citenamefont {Sato}, \citenamefont {Notomi},
  \citenamefont {Shinya}, \citenamefont {Nozaki}, \citenamefont {Taniyama},
  \citenamefont {Hasebe},\ and\ \citenamefont {Kakitsuka}}]{newlaser}%
  \BibitemOpen
  \bibfield  {author} {\bibinfo {author} {\bibfnamefont {S.}~\bibnamefont
  {Matsuo}}, \bibinfo {author} {\bibfnamefont {K.}~\bibnamefont {Takeda}},
  \bibinfo {author} {\bibfnamefont {T.}~\bibnamefont {Sato}}, \bibinfo {author}
  {\bibfnamefont {M.}~\bibnamefont {Notomi}}, \bibinfo {author} {\bibfnamefont
  {A.}~\bibnamefont {Shinya}}, \bibinfo {author} {\bibfnamefont
  {K.}~\bibnamefont {Nozaki}}, \bibinfo {author} {\bibfnamefont
  {H.}~\bibnamefont {Taniyama}}, \bibinfo {author} {\bibfnamefont
  {K.}~\bibnamefont {Hasebe}}, \ and\ \bibinfo {author} {\bibfnamefont
  {T.}~\bibnamefont {Kakitsuka}},\ }\bibfield  {title} {\enquote {\bibinfo
  {title} {Room-temperature continuous-wave operation of lateral current
  injection wavelength-scale embedded active-region photonic-crystal laser},}\
  }\href {\doibase 10.1364/OE.20.003773} {\bibfield  {journal} {\bibinfo
  {journal} {Opt. Express}\ }\textbf {\bibinfo {volume} {20}},\ \bibinfo
  {pages} {3773--3780} (\bibinfo {year} {2012})}\BibitemShut {NoStop}%
\bibitem [{\citenamefont {Ates}\ \emph {et~al.}(2012)\citenamefont {Ates},
  \citenamefont {Sapienza}, \citenamefont {Davanco}, \citenamefont {Badolato},\
  and\ \citenamefont {Srinivasan}}]{quantumdot}%
  \BibitemOpen
  \bibfield  {author} {\bibinfo {author} {\bibfnamefont {S.}~\bibnamefont
  {Ates}}, \bibinfo {author} {\bibfnamefont {L.}~\bibnamefont {Sapienza}},
  \bibinfo {author} {\bibfnamefont {M.}~\bibnamefont {Davanco}}, \bibinfo
  {author} {\bibfnamefont {A.}~\bibnamefont {Badolato}}, \ and\ \bibinfo
  {author} {\bibfnamefont {K.}~\bibnamefont {Srinivasan}},\ }\bibfield  {title}
  {\enquote {\bibinfo {title} {Bright single-photon emission from a quantum dot
  in a circular bragg grating microcavity},}\ }\href {\doibase
  10.1109/JSTQE.2012.2193877} {\bibfield  {journal} {\bibinfo  {journal} {IEEE
  Journal of Selected Topics in Quantum Electronics}\ }\textbf {\bibinfo
  {volume} {18}},\ \bibinfo {pages} {1711--1721} (\bibinfo {year}
  {2012})}\BibitemShut {NoStop}%
\bibitem [{\citenamefont {Zhang}\ and\ \citenamefont {Qiu}(2004)}]{H0cavity}%
  \BibitemOpen
  \bibfield  {author} {\bibinfo {author} {\bibfnamefont {Z.}~\bibnamefont
  {Zhang}}\ and\ \bibinfo {author} {\bibfnamefont {M.}~\bibnamefont {Qiu}},\
  }\bibfield  {title} {\enquote {\bibinfo {title} {Small-volume
  waveguide-section high q microcavities in 2d photonic crystal slabs},}\
  }\href@noop {} {\bibfield  {journal} {\bibinfo  {journal} {Optics express}\
  }\textbf {\bibinfo {volume} {12}},\ \bibinfo {pages} {3988--3995} (\bibinfo
  {year} {2004})}\BibitemShut {NoStop}%
\end{thebibliography}%

\end{document}